\documentclass[10pt,a4paper,abstracton,numbers=noenddot]{scrartcl}
\usepackage[utf8]{inputenc}
\setkomafont{captionlabel}{\rmfamily \bfseries}
\setkomafont{caption}{\rmfamily \small}
\setkomafont{disposition}{\rmfamily \bfseries}
\setcapindent{0em}
\usepackage{color,textcomp,booktabs}
\usepackage{amssymb,amsfonts,amsmath}
\usepackage{braket}
\usepackage[sort&compress]{natbib}
\usepackage[bookmarks=false]{hyperref}
\usepackage{graphicx}
\usepackage{geometry}
\usepackage{bbm}
\usepackage{IEEEtrantools}
\DeclareMathOperator{\tr}{tr}
\allowdisplaybreaks[1]
\def\biggg#1{{\hbox{$\left#1\vbox to25.0pt{}\right.$}}}

\def\Biggg#1{{\hbox{$\left#1\vbox to30.0pt{}\right.$}}}

\title{{\Large Functional Renormalization Group Study of the
Chiral Phase Transition Including Vector and Axial-vector Mesons}}
\author{{\normalsize J\"urgen Eser}\\
{\normalsize \texttt{eser@th.physik.uni-frankfurt.de}}\\
{\normalsize Mara Grahl}\\
{\normalsize \texttt{grahl@th.physik.uni-frankfurt.de}}\\
{\normalsize Dirk H.\ Rischke}\\
{\normalsize \texttt{drischke@th.physik.uni-frankfurt.de}}\\
{\normalsize Johann Wolfgang Goethe-Universit\"at}\\
{\normalsize Max-von-Laue-Str.\ 1, D-60438 Frankfurt am Main, Germany}
}

\begin{document}

\maketitle

\begin{abstract}
\noindent The transition in quantum chromodynamics (QCD) from hadronic matter 
to the quark-gluon plasma (QGP) at high temperatures
and/or net-baryon densities is associated with the restoration
of chiral symmetry and can be investigated in the laboratory via heavy-ion
collisions. We study this chiral transition within
the functional renormalization group (FRG) approach applied to the
two-flavor version of the extended Linear Sigma Model (eLSM).
The eLSM is an effective model for the strong interaction and features besides
scalar and pseudoscalar degrees of freedom also vector and axial-vector mesons.
We discuss the impact of the quark masses and the axial anomaly on the order
of the chiral transition. We also confirm the degeneracy of the masses
of chiral partners above the transition temperature. 
We find that the mass of the $a_{1}$ meson ($\rho$ meson) decreases (increases) 
towards the chiral transition.
\end{abstract}

\section{Introduction}

QCD is the fundamental theory of the strong interaction. For massless quarks, the 
QCD Lagrangian has a global $U(N_{f})_{R}\times U(N_{f})_{L}\cong 
SU(N_{f})_{V}\times SU(N_{f})_{A}\times U(1)_{V}\times U(1)_{A}$ symmetry, 
where $N_{f}$ denotes the number of quark flavors. The $U(1)_V$ symmetry corresponds
to baryon-number conservation, which is always respected. At the quantum level, 
$U(1)_{A}$ is broken to $\mathbb{Z}(N_{f})_{A}$, a phenomenon which is referred to as 
the axial anomaly \cite{Vafa1984}. In the QCD vacuum, the remaining 
$SU(N_{f})_{V}\times SU(N_{f})_{A}$ symmetry, termed ``chiral symmetry'' in the following, is 
further spontaneously broken to $SU(N_{f})_{V}$ by a
non-vanishing quark condensate $\braket{\bar{q}q}$, inducing $N_{f}^{2}-1$ Goldstone bosons 
\cite{Nambu1961a,Nambu1961b,Butti2003}. For nonzero and degenerate 
quark masses, the chiral symmetry is also explicitly broken to $SU(N_{f})_{V}$.

At high temperatures and/or net-baryon number densities, the quark condensate 
melts and chiral symmetry is effectively restored. This chiral transition is commonly 
associated with the so-called QCD transition between a hadronic phase and the QGP. 
The QGP state has existed during the early stages of the universe.
Experiments at accelerator facilities, such as SPS and LHC at CERN, RHIC at BNL, 
or SIS--100/300 at the FAIR project in Darmstadt, aim to explore the
QGP via heavy-ion collisions \cite{Braun-Munzinger2003}. 
Above the chiral transition, the masses of chiral partners, such as the sigma and the pion
or the $\rho$ and the $a_{1}$ meson, become degenerate \cite{Pisarski1995}. 
In particular, dropping $\rho$ and $a_{1}$ 
meson masses were suggested as signatures for chiral symmetry restoration 
\cite{Brown1991,Harada2006,Friman2011}. The change of 
the spectral properties of the $\rho$ meson during the chiral transition 
could, e.g., be detected via its decay
into dileptons \cite{Rapp2000}. Modifications of the dilepton spectrum have been observed 
in Pb+Pb \cite{Agakichiev2005} and In+In \cite{Damjanovic2007} 
collisions. Concerning the axial anomaly, it was claimed in
Ref.\ \cite{Vertesi2011} that data for Au+Au collisions at RHIC energies show a 
reduction of the $\eta'$ meson mass, which was interpreted as a precursor for an effective 
restoration of the $U(1)_{A}$ symmetry \cite{Kapusta1996}.

In general, the order of the chiral transition depends on the number $N_f$ of quark flavors 
and their masses \cite{Fukushima2011a}. Furthermore, for $N_f =2$, the order
depends on whether the $U(1)_{A}$ symmetry is restored
prior to the restoration of chiral symmetry \cite{Gross1981}, 
or whether it remains broken by the axial anomaly.  
For vanishing quark masses, Pisarski and Wilczek \cite{Pisarski1984} 
have argued that, in the absence of the $U(1)_{A}$ anomaly, the chiral phase transition for 
$N_{f}\geq 2$ is of first order and, in the presence of the $U(1)_{A}$ anomaly, can be 
of second order for $N_{f} = 2$. For an infinite anomaly strength and if the transition
is of second order, it definitely falls into the 
$O(4)$-universality class [as originally conjectured by the authors of 
Ref.~\cite{Pisarski1984}]. 
Recently, however, it was argued that the chiral transition can be of second order 
for $N_{f}=2$, no matter whether the axial anomaly is present \cite{Grahl2013}
or not \cite{Pelissetto2013,Grahl2014a}, only the associated universality class differs: 
it was conjectured that the second-order 
transition can be in the $SU(2)_{A}\times U(2)_{V}$ class for finite anomaly
strength, and in the $U(2)_{A}\times U(2)_{V}$ class
for zero anomaly strength, respectively.
For nonzero quark masses, the second-order transition is smeared out into a crossover. 

The chiral transition has been extensively investigated from first principles
via lattice QCD \cite{Bernard2005,Cheng2006,Aoki2006a,Maezawa2007,Cheng2008,
Aoki2009,Bazavov2009,Cheng2010,Bornyakov2010}. The
critical temperature for vanishing quark-chemical potential, $\mu=0$, was
estimated to be $\approx 160$ MeV 
\cite{Bazavov2010,Borsanyi2010,Bornyakov2010,Bazavov2012b,Endrodi2014}. 
Lattice-QCD calculations do not agree upon whether
the  $U(1)_{A}$ symmetry has already been restored at the
chiral transition temperature \cite{Aoki2012,Cossu2013}
or not  \cite{Sharma2013,Bhattacharya2014,Dick:2015twa}.
Unfortunately, perturbative methods fail to achieve 
reliable results in the low-energy range, where the gauge 
coupling is strong. Besides, they are plagued by severe infrared divergences. 
Therefore, non-perturbative continuum methods, such as the functional renormalization group 
(FRG) \cite{Wetterich1993,Morris1994,Pawlowski2007,Kopietz2010}, are
complementary to lattice QCD and hence indispensable for exploring the nature of QCD matter.

FRG studies of the chiral transition using effective models for QCD
have a long tradition \cite{Berges2002}: a quark-meson model with $U(2)_{R}\times U(2)_{L}$ symmetry 
has been investigated beyond the local potential 
approximation (LPA) in Ref.~\cite{Jungnickel1996,Schaefer2008}. An $O(N)$-symmetric quark-meson model
was studied in Ref.~\cite{Berges:1997eu}.
The phase structures of the chiral quark-meson model and 
the Polyakov-quark-meson(-diquark) model for the two lightest flavors as well as for $2+1$ flavors
were determined in Refs.\ \cite{Schaefer2005,Schaefer2007a,Schaefer2007b,
Herbst:2010rf,Skokov:2010wb,Braun:2007bx,Strodthoff:2011tz,Herbst:2013ail,Herbst:2013ufa,Tripolt:2013jra,Strodthoff:2013cua}.
$U(1)_A$-violating terms in the FRG were introduced in Ref.\ \cite{Pawlowski:1996ch}.
For massless up and down quarks and a physical strange quark, 
the chiral transition was shown to be of first/second order without/with the 
$U(1)_{A}$ anomaly in the framework of a quark-meson model \cite{Schaefer:2013isa,Mitter2014}.
Also non-vanishing external magnetic fields were studied in
such a model \cite{Andersen:2013swa,Kamikado2014}. 
Recent insights into second-order and 
fluctuation-induced first-order phase transitions in a $U(2)_{R}\times U(2)_{L}$ symmetric
model with scalar and pseudoscalar mesons were provided e.g.\ in Refs.\
\cite{Fukushima2011b,Fejos2014}. 
The influence of the $U(1)_{A}$ anomaly in a purely mesonic model
was studied in Ref.~\cite{Grahl2013}. The infrared-stable 
$U(2)_{V}\times U(2)_{A}$ fixed point detected in Ref.~\cite{Pelissetto2013} was also 
found with the help of the FRG technique, but the 
subsequent stability analysis remained inconclusive \cite{Grahl2014a}. 
FRG studies based on QCD degrees of freedom were presented in Refs.\ 
\cite{Pawlowski2007,Braun2010b,Braun2011,Fister2013,Braun2014,Mitter:2014wpa}. 

Vector-meson fields were introduced into a chiral effective nucleon-meson
Lagrangian in Refs.\ \cite{Drews2013,Drews2014a,Drews2014b,Drews2014c,Drews2015}
and the phase diagram at nonvanishing net-baryon density was studied within
the FRG. However, the vector-meson degrees of freedom were only considered 
to be background fields, i.e., their fluctuations were neglected in the FRG flow.  
A first FRG study of (axial-)vector mesons in QCD at $T=\mu=0$ was performed in 
Ref.~\cite{Rennecke2015}, revealing that the hadronic 
contributions to the flow in vacuum are dominated by the sigma meson and the pions.

A study of the chiral transition at nonzero temperature within the FRG including 
vector and axial-vector mesons is, however, still missing. In this work, we fill this gap
by applying the FRG to a purely mesonic model.
We ascertain the mass degeneracy of chiral 
partners above the transition and identify its order.
We consider the so-called eLSM  
\cite{Parganlija2010,Parganlija2012,Parganlija2013} as effective model for the 
strong interaction at nonzero $T$ and for vanishing $\mu$. This model
has been shown to reproduce hadronic vacuum properties such as masses and
decay widths to a surprising degree of accuracy \cite{Parganlija2013}. 
Furthermore, it has the same low-energy limit as QCD \cite{Divotgey2015}.
We study this model in a limit
where it resembles the time-honored Sakurai model \cite{Sakurai1960,Strueber2008}. 

This paper is organized as follows: in Sec.~\ref{sec:LSM} we introduce 
the two-flavor version of the eLSM. The subsequent part 
(Sec.~\ref{sec:FRG}) discusses the FRG formalism 
to describe fluctuations of spin-zero and spin-one mesons.
The numerical results for vanishing quark masses in the absence (or presence) of the 
axial anomaly are presented in Sec.~\ref{sec:u2u2} (\ref{sec:anomaly}). 
The case of explicit breaking of chiral symmetry by nonzero quark masses is
discussed in Sec.~\ref{sec:esb}. Section \ref{sec:discussion} concludes this work with 
a summary of our results and an outlook.

We use natural units $\hbar=c=k_{B}=1$ and work in a finite $(3+1)$-dimensional 
Euclidean spacetime volume $V\times (0,1/T]$ 
at nonzero temperature $T$ with periodic boundary conditions and, consequently, a 
discrete momentum spectrum: $q=(\omega_{n},\vec{q})$, where the Matsubara frequencies 
for bosonic fields are given by  $\omega_{n}=2n\pi T$. We use a shorthand notation for 
spacetime integrations:
\begin{equation}
	\int_{\mathcal{V}}\mathrm{d}^{3+1}x
	=\int^{1/T}_{0}\mathrm{d}\tau\int_{V}{\mathrm{d}^{3}x}=\int_{x}\ ,
\end{equation}
with $\mathcal{V}=V/T$.
We employ Einstein's summation convention, i.e., indices appearing twice are summed over.
If these indices are Lorentz indices, $\mu = 0,1,2,3$, it does not matter whether they appear
as co- or contravariant indices, because we work in Euclidean spacetime. We shall always use
covariant Lorentz indices.

\section{Methods}

\subsection{Extended Linear Sigma Model}\label{sec:LSM}
At low energies, quarks and gluons are confined inside hadrons, which are
thus the effective degrees of freedom. An effective theory for hadrons must
incorporate the chiral symmetry of QCD, as well as its spontaneous breaking. 
We work with a mesonic linear sigma model \cite{Schwinger1957,Gell-Mann1960} 
as an effective implementation of the strong interaction, which, 
besides scalar and pseudoscalar mesons, also includes vector and axial-vector mesons 
\cite{Gasiorowicz1969,Boguta1983,Ko1994,Pisarski1995,Parganlija2010}:
the so-called eLSM \cite{Parganlija2012}. 
The scalar and pseudoscalar fields are the real and imaginary parts of
a complex $N_{f}\times 
N_{f}$ matrix $\Sigma$ that lives in the $[N_f^*,N_f]$ representation
of the group $U(N_f)_R \times U(N_f)_L$. Under transformations of this group, 
$\Sigma$ behaves as follows:
\begin{equation}
	\Sigma\rightarrow U_{R}^{\dagger}\Sigma U_{L}\ ,\label{eqn:trans0}
\end{equation}
where the group elements $U_{R,L}$ are unitary matrices. In terms of hadronic fields, $
\Sigma=(\sigma_{a}+i\pi_{a})t_{a}$, with the
generators $t_{a}$ of $U(2)$ in the fundamental representation ($\mathrm{tr}[t_{a}t_{b}] =\delta_{ab}/
2$). Here, $\sigma_{a}$ and $\pi_{a}$ represent scalar and 
pseudoscalar degrees of freedom, respectively. Analogously, we define right- and left-handed 
fields for (axial\hbox{-})\-vector mesons 
(parametrized by axial-vector fields $A_{a,\mu}$ and vector fields $V_{a,\mu}$): 
$R_{\mu}=(V_{a,\mu}+A_{a,\mu})t_{a}$, $L_{\mu}
=(V_{a,\mu}-A_{a,\mu})t_{a}$. They transform as:
\begin{equation}
	R_{\mu}\rightarrow U_{R}^{\dagger}R_{\mu}U_{R}\ ,\quad L_{\mu}\rightarrow U_{L}^{\dagger}
	L_{\mu}U_{L}\ .\label{eqn:trans1}
\end{equation}
The globally chirally symmetric Lagrangian is given by \cite{Parganlija2012}:
\begin{IEEEeqnarray}{rCl}
	\mathcal{L} & = & \tr\left[\left(D_{\mu}\Sigma\right)^{\dagger} D_{\mu}\Sigma\right]
	+m_{0}^{2} 
	\tr\left(\Sigma^{\dagger}\Sigma\right)+\lambda_{1}\left[\tr\left(\Sigma^{\dagger}\Sigma
	\right)\right]^{2}+\lambda_{2}
	\tr\left[\left(\Sigma^{\dagger}\Sigma\right)^{2}\right]+\frac{1}{4}\tr\left(L_{\mu\nu}
	^{2}+R_{\mu\nu}^{2}\right)\nonumber\\
	& & +\tr\left[\left(\frac{m_{1}^{2}}{2}+\Delta\right)\left(L_{\mu}^{2}+R_{\mu}^{2}\right)
	\right]-\tr\left[H\left(\Sigma 
	+\Sigma^{\dagger}\right)\right]-c_{A}\left(\det\,\Sigma+\det
	\, \Sigma^{\dagger}\right)\nonumber\\
	& & -ig_{2}\left(\tr\left\lbrace L_{\mu\nu}\left[L_{\mu},L_{\nu}\right]\right\rbrace 
	+\tr\left\lbrace R_{\mu\nu}\left[R_{\mu},R_{\nu}\right]\right\rbrace\right)
	+\frac{h_{1}}{2}\tr\left(\Sigma^{\dagger}\Sigma\right)\tr\left(L_{\mu}^{2}+R_{\mu}^{2}
	\right)\nonumber\\
	& & +h_{2}\tr\left(\left|R_{\mu}\Sigma\right|^{2}+\left|\Sigma L_{\mu}\right|^{2}\right)
	+2h_{3}\tr\left(\Sigma L_{\mu}\Sigma^{\dagger}R_{\mu}\right)-g_{3}\left[\tr\left(L_{\mu}
	L_{\nu}L_{\mu}L_{\nu}\right)
	+\tr\left(R_{\mu}R_{\nu}R_{\mu}R_{\nu}\right)\right]\nonumber\\
	& & -g_{4}\left[\tr\left(L_{\mu}L_{\mu}L_{\nu}L_{\nu}\right)+\tr\left(R_{\mu}R_{\mu}
	R_{\nu}R_{\nu}\right)\right]
	-g_{5}\tr\left(L_{\mu}L_{\mu}\right)\tr\left(R_{\nu}R_{\nu}\right)\nonumber\\
	& & -g_{6}\left[\tr\left(L_{\mu}L_{\mu}\right)\tr\left(L_{\nu}L_{\nu}\right)
	+\tr\left(R_{\mu}R_{\mu}\right)\tr\left(R_{\nu}R_{\nu}\right)\right]\ ,\label{eqn:Lglobal}
\end{IEEEeqnarray}
with the covariant derivative $D_{\mu}\Sigma=\partial_{\mu}\Sigma+ig_{1}(\Sigma L_{\mu}-
R_{\mu}\Sigma)$ and the right-/left-handed field-strength 
tensors $R_{\mu\nu}=\partial_{\mu}R_{\nu}-\partial_{\nu}R_{\mu}$ and $L_{\mu\nu}=
\partial_{\mu}L_{\nu}-\partial_{\nu}L_{\mu}$. The term
$\det\,\Sigma+\det\, \Sigma^{\dagger}$ accounts for the $U(1)_{A}$ anomaly 
by breaking $U(N_f)_R \times U(N_f)_L$ to $SU(N_{f})_{V}\times 
SU(N_{f})_{A}\times U(1)_{V}$. Its strength is determined by the coupling constant $c_{A}$. 
The flavor-diagonal terms $\tr\left[H\left(\Sigma 
+\Sigma^{\dagger}\right)\right]$ and $\tr\left[\Delta\left(L_{\mu}^{2}+R_{\mu}^{2}\right)
\right]$ correspond to explicit symmetry breaking
(ESB) in the (pseudo\hbox{-})\-scalar and (axial\hbox{-})\-vector sector, respectively: 
\begin{equation}
	H=\sum_{i=1}^{N_{f}}h_{0}^{i^2-1}t_{i^2-1}\ ,\quad
	\Delta=\mathrm{diag}\left[\delta_{1},\delta_{2},\ldots,\delta_{N_{f}}\right]\propto
	\mathrm{diag}
	\left[m^{2}_{u},m^{2}_{d},\ldots,m^{2}_{N_{f}}\right]\ .
\end{equation}
For nonzero and degenerate quark masses $m^{2}_{u}=m^{2}_{d}=\ldots$ ($h_{0}^{0}\neq 0$, 
while all other $h_{0}^{i^2-1}$ vanish; 
$\Delta\propto\mathbbmss{1}$ has no further impact), 
these terms break the $U(N_{f})_{R}\times U(N_{f})_{L}$ symmetry to $U(N_{f})_{V}$.

It should be mentioned that there is a second way of introducing spin-one degrees of freedom 
to this effective theory. Within the gauged linear 
sigma model (gLSM) \cite{Gasiorowicz1969,Strueber2008}, 
(axial\hbox{-})\-vector mesons are treated as massive Yang-Mills fields, accounting for
the phenomenon of vector-meson dominance \cite{Sakurai1960,Urban2002}. This model is 
constructed by requiring local 
$U(N_{f})_{R}\times U(N_{f})_{L}$ symmetry. The gauge principle calls for a universal 
coupling of right- and left-handed vector fields to 
(pseudo\hbox{-})\-scalars as well as among spin-one fields themselves. But due to the nonzero 
mass of the ``gauge bosons'', of course, the gLSM 
is not a true gauge theory and the local invariance is already broken down to a global one. 
Since chiral symmetry is of global nature in QCD 
anyway, it seems to be logical to work with the Lagrangian (\ref{eqn:Lglobal}). Moreover, the 
``local'' version does not reproduce the correct 
phenomenology of $\rho$ and $a_{1}$ mesons \cite{Pisarski1995,Urban2002}, a problem which is 
solved by the globally symmetric eLSM \cite{Parganlija2013}.

In this study, we restrict ourselves to the isospin-symmetric two-flavor case, i.e.,
up and down quarks have the same mass. Hence we are dealing with scalar fields 
$(\sigma,\vec{a}_{0})$, pseudoscalars $(\eta,\vec{\pi})$, and with $(f_{1},\vec{a}_{1})$ as 
well as $(\omega,\vec{\rho})$ in the 
(axial\hbox{-})\-vector mesonic sector. The fields
$\sigma$, $\eta$, $\omega$, and $f_{1}$ are $SU(2)$-singlet states, whereas the others 
form isospin triplets. The field $\eta$ does not correspond to the physical $\eta$/$\eta'$
mesons, which are mixtures of $\bar{n}n$ and $\bar{s}s$ ($n=u,d$ stands for the nonstrange
up and down quarks, $s$ for the strange quark). 
As a first step in applying the FRG to the eLSM, we want to keep 
things simple and set the constants $g_{2},g_{3},
\ldots ,g_{6}$ as well as $h_{1},h_{2},h_{3}$ to zero. The above defined complex $2\times 2$-
matrices explicitly read:
\begin{IEEEeqnarray}{rCl}
	\Sigma & = & (\sigma+i\eta)t_{0}+(\vec{a}_{0}+i\vec{\pi})\cdot\vec{t}\ ,\\
	R_{\mu} & = & (\omega_{\mu}+f_{1\mu})t_{0}+(\vec{\rho}_{\mu}+\vec{a}_{1\mu})\cdot\vec{t}\ 
	,\\
	L_{\mu} & = & (\omega_{\mu}-f_{1\mu})t_{0}+(\vec{\rho}_{\mu}-\vec{a}_{1\mu})\cdot\vec{t}\ 
	,
\end{IEEEeqnarray}
and the different parts of $\mathcal{L}$ can be expressed as:
\begin{IEEEeqnarray}{rCl}
	\tr\left[\left(D_{\mu}\Sigma\right)^{\dagger} D_{\mu}\Sigma\right] & = & \frac{1}{2}
	\left[\partial_{\mu}\sigma
	+g_{1}\left(\eta f_{1\mu}+\vec{\pi}\cdot\vec{a}_{1\mu}\right)\right]^{2}+\frac{1}{2}
	\left[\partial_{\mu}\eta-g_{1}
	\left(\sigma f_{1\mu}+\vec{a}_{0}\cdot\vec{a}_{1\mu}\right)\right]^{2}\nonumber\\
	& & +\frac{1}{2}\left[\partial_{\mu}\vec{a}_{0}+g_{1}\left(\vec{\rho}_{\mu}\times\vec{a}
	_{0}+\eta \vec{a}_{1\mu}
	+\vec{\pi}f_{1\mu}\right)\right]^{2}\nonumber\\
	& & +\frac{1}{2}\left[\partial_{\mu}\vec{\pi}-g_{1}\left(\vec{\pi}\times\vec{\rho}_{\mu}+
	\sigma \vec{a}_{1\mu}
	+\vec{a}_{0}f_{1\mu}\right)\right]^{2}\ ,\label{eqn:L1}\\
	m_{0}^{2}\tr\left(\Sigma^{\dagger}\Sigma\right) & = & \frac{m_{0}^{2}}{2}\left(\sigma^{2}
	+\vec{a_{0}}^{2}+\eta^{2}
	+\vec{\pi}^{2}\right)\ ,\\
	\lambda_{1}\left[\tr\left(\Sigma^{\dagger}\Sigma\right)\right]^{2} & = & 
	\frac{\lambda_{1}}{4}\left(\sigma^{2}
	+\vec{a_{0}}^{2}+\eta^{2}+\vec{\pi}^{2}\right)^{2}\ ,\\
	\lambda_{2}\tr\left[\left(\Sigma^{\dagger}\Sigma\right)^{2}\right] & = & 
	\frac{\lambda_{2}}{4}\bigg\lbrace\frac{1}{2}
	\left(\sigma^{2}+\vec{a_{0}}^{2}+\eta^{2}+\vec{\pi}^{2}\right)^{2}\nonumber\\
	& & +2\left[\left(\sigma^2+\vec{\pi}^2\right)\left(\eta^2+\vec{a}_{0}^2\right)-
	\left(\sigma\eta-\vec{\pi}\cdot\vec{a}_{0}
	\right)^{2}\right]\bigg\rbrace\ ,\\
	\frac{1}{4}\tr\left(L_{\mu\nu}^{2}+R_{\mu\nu}^{2}\right) & = & \frac{1}{4}
	\left(\partial_{\mu}\omega_{\nu}
	-\partial_{\nu}\omega_{\mu}\right)^{2}+\frac{1}{4}\left(\partial_{\mu}\vec{\rho}_{\nu}
	-\partial_{\nu}\vec{\rho}_{\mu}\right)^{2}\nonumber\\
	& & +\frac{1}{4}\left(\partial_{\mu}f_{1\nu}-\partial_{\nu}f_{1\mu}\right)^{2}
	+\frac{1}{4}\left(\partial_{\mu}\vec{a}_{1\nu}-\partial_{\nu}\vec{a}_{1\mu}\right)^{2}\ ,
	\\
	\frac{m_{1}^{2}}{2}\tr\left(L_{\mu}^{2}+R_{\mu}^{2}\right) & = & \frac{m_{1}^{2}}{2}
	\left(f_{1\mu}^{2}+\vec{a}_{1\mu}^{2}+\omega_{\mu}^{2}+\vec{\rho}_{\mu}^{2}\right)\ ,\\
	\tr\left[H\left(\Sigma+\Sigma^{\dagger}\right)\right] & = & h_{0}^{0}\sigma\ ,\\
	c_{A}\left(\det\, \Sigma+\det\, \Sigma^{\dagger} \right) & = & 
	\frac{c_{A}}{2}\left(\sigma^{2}-\vec{a_{0}}^{2}-\eta^{2}+\vec{\pi}^{2}\right)\ .
\end{IEEEeqnarray}
Obviously, the $\omega$ meson completely decouples from any interactions. This would not be 
the case had we included other
terms from Eq.~(\ref{eqn:Lglobal}). 

In the low-temperature broken phase, the isoscalar $\sigma$ field acquires a non-vanishing 
expectation value $\braket{\sigma}_0=\braket{\bar{q}q}=\mathrm{const.}\neq 0$ (here, the 
angular brackets are the notation for expectation values and the subscript 0 denotes
the absence of external sources). 
Therefore, one has to consider fluctuations around the physical ground 
state and thus a shift of the $\sigma$ field by its expecation value: $\sigma\rightarrow
\braket{\sigma}_0+\sigma$. The expectation value $\braket{\sigma}_0$ acts as the 
order parameter for the chiral phase transition. After accounting for this shift, 
integration by parts then gives rise to the 
bilinear terms $g_{1}\braket{\sigma}_0\eta\, \partial_{\mu}f_{1\mu}$ and $g_{1}\braket{\sigma}_0
\vec{\pi}\cdot \partial_{\mu}\vec{a}_{1\mu}$. 
They represent the so-called $\pi$-$a_{1}$- and $\eta$-$f_{1}$-mixing, leading to nondiagonal 
elements in the scattering matrix. Usually these 
terms are eliminated. Following Ref.~\cite{Strueber2008}, this is done by shifting the 
axial-vector fields: $f_{1\mu}\rightarrow 
f_{1\mu}+w\partial_{\mu}\eta$ and $\vec{a}_{1\mu}\rightarrow \vec{a}_{1\mu}+w\partial_{\mu}
\vec{\pi}$ with 
$w=g_{1}\braket{\sigma}_0/[m_{1}^{2}+(g_{1}\braket{\sigma}_0)^{2}]$. In 
turn, the axial-vector fields become explicitly RG-scale dependent (through the dependence of
$w$ on $\braket{\sigma}_0$) and the 
pseudoscalar states need to be renormalized: $\pi_{a}\rightarrow \sqrt{Z_{\pi}}\, \pi_{a}$, 
$Z_{\pi}=1+(g_{1}\braket{\sigma}_0)^{2}/m_{1}^{2}$. This provides the canonical 
normalization of all one-meson states, such that their Fourier 
components can be interpreted as creation and annihilation operators in the process of 
quantization \cite{Gasiorowicz1969}. For a precise 
discussion of the $\sigma$ shift and its implications on the FRG flow we refer to 
Ref.~\cite{Rennecke2015}. Instead of redefining the $a_{1}$ 
and $f_{1}$ fields, one may also work with nondiagonal propagators as performed in 
Ref.~\cite{Urban2002}.

The vacuum expectation value $\braket{\sigma}_{0}$ is the minimum of 
the classical potential energy density 
$V(\braket{\sigma})$:
\begin{equation}
	V(\braket{\sigma})=\frac{1}{2}\left(m_{0}^{2}-c_{A}\right)\braket{\sigma}^{2}+\frac{1}{4}
	\left(\lambda_{1}
	+\frac{\lambda_{2}}{2}\right)\braket{\sigma}^{4}-h_{0}^{0}\braket{\sigma}\ ,
	\quad\left.\frac{\mathrm{d}V}{\mathrm{d}\braket{\sigma}}\right|_{\braket{\sigma}=
	\braket{\sigma}_{0}}=0\ .
\end{equation}
The wave-function renormalization $Z_{\pi}$ is related to $\braket{\sigma}_{0}$ and to the 
masses of the $a_{1}$ and $\rho$ mesons by:
\begin{equation}
	\braket{\sigma}_{0}=\sqrt{Z_{\pi}}\, f_{\pi}\equiv f_{\pi}\frac{m_{a_{1}}}{m_{\rho}}\ ,
\end{equation}
where $f_{\pi}\simeq 93$ MeV is the pion decay constant. The KSFR 
(Kawaraba\-yashi-Su\-zuki-Fay\-yazud\-din-Riazud\-din) relation 
\cite{Kawarabayashi1966,Riazuddin1966} predicts that $Z_{\pi}=2$. This slightly 
differs from the value of $Z_{\pi} \simeq (m_{a_1}/m_\rho)^2 \simeq 2.552$ 
quoted by the particle data group \cite{Olive2014}. 

The identification of the mesonic fields with measured resonances listed in Ref.~\cite{Olive2014} 
is partly straightforward: the pions and the 
$\eta$ (as the pure $\bar{n}n$ state arising from unmixing the physical $\eta$ and $\eta'$) 
have a mass around 140 MeV and 700 MeV, respectively. The vector fields $\omega$ and $\rho$ 
represent the $\omega(782)$ and $\rho(770)$ resonances. The axial vectors $f_1$ and
$a_1$ correspond to the $f_{1}(1285)$ and $a_{1}(1260)$. For the $\sigma$ and the 
$a_{0}$ fields, however, it is controversial whether they should describe $\lbrace f_{0}
(500),a_{0}(980)\rbrace$ or $\lbrace f_{0}(1370),a_{0}
(1450)\rbrace$. It was argued in 
Refs.~\cite{Parganlija2010,Parganlija2012,Parganlija2013} that the latter option might be favored.

\subsection{Functional renormalization group}\label{sec:FRG}

The Wilsonian renormalization group (RG) performs the mode integration of 
(quantum-)statistical fluctuations from the ultraviolet (UV) to the 
infrared (IR) in a stepwise manner, i.e., it successively takes momentum-shell 
by momentum-shell into account 
\cite{Wilson1969,Wilson1971b,Wilson1971c,Wilson1972b,Wilson1972a,Wilson1974,Wilson1975}. The 
FRG is an implementation of this procedure which 
allows us to non-perturbatively formulate quantum field theories in terms of a differential 
equation. This flow equation dictates the scale ($k$-)
dependence of the effective average action $\Gamma_{k}$, which interpolates between the bare 
interactions at some UV cutoff scale $k_{\mathrm{UV}}=\Lambda$ and 
the macroscopic physics including all fluctuations in the IR, $k_{\mathrm{IR}} =0$. 
A $k$-dependent term $\Delta S_{k}$ is added to the action $S$ in order to provide 
an effective cutoff at momenta $q^{2}\simeq k^{2}$, such that only modes with 
$q^{2}\gtrsim k^{2}$ are integrated out in the RG flow. The term $\Delta S_{k}$ 
regulates the scale evolution of $\Gamma_{k}$ in such a way that the full effective action $
\Gamma\equiv \Gamma_{k\rightarrow 0}$ is obtained in 
the IR limit. The effective action $\Gamma$ is the generating functional 
of one-particle irreducible vertex diagrams of the theory. 
For $k\rightarrow \Lambda$, in contrast, the classical 
action is recovered: $\Gamma_{k\rightarrow\Lambda}= S$.

Our investigations focus on the theory defined by the Lagrangian (\ref{eqn:Lglobal}).
In order to simplify the following discussion,
we denote (pseudo-)scalar fields by $\varphi_i$ and (axial-)vector fields
by $A_{i,\mu}$. The field-strength tensors of the latter
are denoted as $F_{i,\mu\nu}=\partial_{\mu}A_{i,\nu}-
\partial_{\nu}A_{i,\mu}$. The fields $\varphi_i$ and $A_{i,\mu}$ are subject to thermal (for $T>0$)
as well as to quantum fluctuations (also at $T=0$).
Following the discussion in Ref.~\cite{Rischke1994}, we apply 
Stueckelberg's Lagrangian 
\cite{Itzykson1985,Ruegg2004} with coupling $\lambda_{\mathrm{St}}$ to derive the FRG flow 
equation:
\begin{equation}
	S=\int_{x}\mathcal{L}\ \rightarrow\ \int_{x}\left(\mathcal{L}+
	\frac{\lambda_{\mathrm{St}}}{2}\partial_{\mu}
	A_{i,\mu}\partial_{\nu}A_{i,\nu}\right)
	\ .\label{eqn:action}
\end{equation}
(Axial\hbox{-})\-vector mesonic fields usually 
have three physical degrees of freedom. The additional term in Eq.~(\ref{eqn:action}), 
however, promotes the unphysical fourth to a 
physical one. Hence, in the following, all vector fields initially have four instead of three 
degrees of freedom. Although not necessary to 
ensure renormalizability \cite{Rennecke2015}, this formalism guarantees that we work with invertible inverse 
tree-level propagators \cite{Strueber2008}. 
Furthermore, this strategy allows to derive the grand canonical partition function in a 
simple manner \cite{Rischke1994}.

As a starting point for deriving the FRG flow equation of the theory at hand, we consider the 
scale-dependent generating functional $W_{k}$ for 
connected Green's functions:
\begin{equation}
	W_{k}\left[J_{i},J_{i,\mu}\right]\equiv\ln Z_{k}\left[J_{i},J_{i,\mu}\right]=\ln\int
	\mathcal{D}\varphi_{i}
	\mathcal{D}A_{i,\mu}\ e^{-S\left[\varphi_{i},A_{i,\mu}\right]-\Delta S_{k}
	\left[\varphi_{i},A_{i,\mu}\right]
	+\int_{x}J_{i}\varphi_{i}+\int_{x}J_{i,\mu}A_{i,\mu}}\ .\label{eqn:Wk}
\end{equation}
As discussed above, we add a regulator term $\Delta S_{k}$ to the action:
\begin{equation}
	\Delta S_{k}\left[\varphi_{i},A_{i,\mu}\right]=\frac{1}{2}\mathcal{V}\sum_{q}
	\left[{\varphi_{i}(-q)R_{k}^{\mathrm{S}}(q)
	\varphi_{i}(q)+A_{i,\mu}(-q)R^{\mathrm{V}}_{k,\mu\nu}(q)A_{i,\nu}(q)}\right]\ ,
	\label{eqn:deltaS}
\end{equation}
which can be interpreted as a momentum-dependent mass term, and we also included sources 
$J_{i}$ and $J_{i,\mu}$ for scalar and vector fields. 
To ensure the required UV/IR limits for the flow of $\Gamma_{k}$, the regulator functions 
$R_{k}^{\mathrm{S}}(q)$ and $R^{\mathrm{V}}_{k,\mu\nu}
(q)$ must fulfill the following relations: $R_{k}^{\mathrm{S}}(q),R^{\mathrm{V}}_{k,\mu\nu}
(q)\rightarrow 0$ for $k\rightarrow0$, as well as 
$R_{k}^{\mathrm{S}}(q),R^{\mathrm{V}}_{k,\mu\nu}(q)\rightarrow\infty$ for $k\rightarrow
\Lambda$. On top of that, the regulators should satisfy: 
$R_{k}^{\mathrm{S}}(q),R^{\mathrm{V}}_{k,\mu\nu}(q)\sim k^{2}$ for $q\rightarrow 0$ and 
$R_{k}^{\mathrm{S}}(q),R^{\mathrm{V}}_{k,\mu\nu}(q)\sim 
0$ for $q\rightarrow\infty$. Apparently, low-energy fluctuations ($q^{2}\ll k^{2}$) are 
effectively separated from the RG integration process by 
giving them an additional ``mass'' $\sim k^{2}$, while fast modes ($q^{2}\gg k^{2}$) are not 
influenced. The effective average action 
$\Gamma_{k}+ \Delta S_k$ is the Legendre transform of $W_{k}$, or in other words:
\begin{equation}
	\Gamma_{k}\left[\phi_{k,i},\mathcal{A}_{k,i,\mu}\right]= 
	\mathcal{V}\sum_{q}\left[J_{i}(-q)\phi_{k,i}(q)+J_{i,\mu}(-q)\mathcal{A}_{k,i,\mu}(q)
	\right]
	-W_{k}\left[J_{i},J_{i,\mu}\right]-\Delta S_{k}\left[\phi_{k,i},\mathcal{A}_{k,i,\mu}
	\right]\ ,\label{eqn:Legendre}
\end{equation}
where $\phi_{k,i}(q)=\braket{\varphi_{i}(q)} \equiv \mathcal{V}^{-1} \delta  W_k/\delta J_i(-q)$ and 
$\mathcal{A}_{k,i,\mu}(q)=\braket{A_{i,\mu}(q)} \equiv  \mathcal{V}^{-1}  \delta W_k/\delta J_{i,\mu}(-q)$
are the expectation values of the fields in the presence of the sources
$J_{i}$ and $J_{i,\mu}$. 
Although the Legendre transform $\Gamma_{k}+ \Delta S_k$  is convex by definition, 
this does not hold for $\Gamma_{k}$ itself, as $\Delta S_{k}$ is not necessarily curved in the same 
way. Exclusively in the case $k\rightarrow 0$, where $\Delta S_k \rightarrow 0$,
$\Gamma_{k\rightarrow 0} = \Gamma$
becomes the true Legendre transform of $W_{k\rightarrow 0}\equiv W$, and thus is 
definitely convex. 

For fixed values of the fields, differentiation of 
Eq.~(\ref{eqn:Legendre}) with respect to $k$ yields the FRG flow equation:
\begin{equation}
	\partial_{k}\Gamma_{k}=\frac{1}{2}\mathcal{V}\sum_{q}\left\lbrace\tr\left[\mathbf{G}
	^{\mathrm{S}}_{k}(q,q) 
	\partial_{k}\mathbf{R}_{k}^{\mathrm{S}}(q)\right]+\tr\left[\mathbf{G}^{\mathrm{V}}_{k,\mu
	\nu}(q,q)\partial_{k} 
	\mathbf{R}^{\mathrm{V}}_{k,\nu\mu}(q)\right]\right\rbrace\ .\label{eqn:floweq}
\end{equation}
Here, $\mathbf{G}^{\mathrm{S}}_{k}$ and $\mathbf{G}^{\mathrm{V}}_{k,\mu\nu}$ denote the full 
propagators for scalar and vector fields. Introducing 
the general field notation $\Phi=(\phi_{k,i},\mathcal{A}_{k,i,\mu})$ and using the fact that 
$\mathcal{V}\mathbf{G}_{k}=\left(\mathbf{\Gamma}^{(2)}_{k}+\mathbf{R}_{k}\right)^{-1}$, 
Eq.~(\ref{eqn:floweq}) simplifies to:
\begin{equation}
	\partial_{k}\Gamma_{k}=\frac{1}{2}\tr\left[\left(\mathbf{\Gamma}^{(2)}_{k}+\mathbf{R}_{k}
	\right)^{-1}\partial_{k}
	\mathbf{R}_{k}\right]\ .\label{eqn:wetterich}
\end{equation}
The momentum summation has been included in the definition of the trace. The
propagators $\mathbf{G}_{k}$ and $
\mathbf{R}_{k}$ are matrix-valued in momentum space and 
in all internal spaces. 

In principle, the FRG equation and the resulting macroscopic physical 
observables should be independent of the form of the
regulators, which are only restricted by the limits discussed above. In this case, all trajectories 
in coupling space predicted by different choices of 
$\mathbf{R}_{k}$ start at the point $\Gamma_{k\rightarrow \Lambda}=S$ and terminate at $
\Gamma_{k\rightarrow 0}=\Gamma$. In practice, however, 
one needs to truncate the infinite hierarchy of flow equations arising from 
Eq.~(\ref{eqn:wetterich}) in order to solve them. 
Indeed, this fact inevitably leads to a regulator-dependent bias, which, fortunately, can be 
minimized by working with the optimized Litim 
regulator \cite{Litim2001}. 

One convenient truncation scheme is the expansion of $\Gamma_{k}$ 
in terms of field derivatives 
\cite{Berges2002,Kopietz2010}. For a purely scalar theory, this would 
read:
\begin{equation}
	\Gamma_{k}[\Phi]=\int_{x}\left[U_{k}(\Phi)+\frac{1}{2}\mathcal{Z}_{k}(\Phi)\partial_{\mu}
	\Phi\cdot\partial_{\mu}\Phi
	+\mathcal{O}(\partial^{4})\right]\ .\label{eqn:derexp}
\end{equation}
$U_{k}$ is the scale-dependent effective potential, $\mathcal{Z}_{k}$ symbolizes the 
wave-function renormalization associated to the scaling of the kinetic part.
The LPA assumes the wave-function renormalization in the derivative expansion (\ref{eqn:derexp}) 
to be field-independent and fixed to its initial value of one, $\mathcal{Z}_{k}(\Phi)=\mathcal{Z}
_{k}=1$, which corresponds to a vanishing anomalous 
dimension. Momentum-dependent interactions are neglected in LPA. 
In the case of the eLSM, the second term in 
Eq.~(\ref{eqn:derexp}) is given by the kinetic terms of spin-zero 
and spin-one fields (modified by Stueckelberg's Lagrangian):
\begin{equation}
	\Gamma_{k}[\phi_{i},\mathcal{A}_{i}]=\int_{x}\left[\frac{1}{2}\partial_{\mu}\phi_{i}
	\partial_{\mu}\phi_{i}
	+\frac{1}{4}\mathcal{F}_{i,\mu\nu}\mathcal{F}_{i,\mu\nu}
	+\frac{\lambda_{\mathrm{St}}}{2}\left(\partial_{\mu}\mathcal{A}_{i,\mu}\right)^{2}
	+U_{k}(\phi_{i},\mathcal{A}_{i})\right]\ ,\label{eqn:derexpLPA}
\end{equation}
$\mathcal{F}_{i,\mu\nu}=\partial_{\mu}\mathcal{A}_{i,\nu}-\partial_{\nu}\mathcal{A}_{i,\mu}$. 
At nonzero temperature, it is technically 
advantageous to employ the three-dimensional version of Litim's optimal regulator $\propto 
(k^{2}-\vec{q}^{\,2})\Theta(k^{2}-\vec{q}^{\,2})$ 
\cite{Litim2006,Blaizot2007}. Remembering the above discussion about their various
limits, the regulating functions are chosen as:
\begin{IEEEeqnarray}{rCl}
	R_{k}^{\mathrm{S}}(q) & = & (k^{2}-\vec{q}^{\,2})\Theta(k^{2}-\vec{q}^{\,2})\ ,
	\label{eqn:RS}\\
	R^{\mathrm{V}}_{k,\mu\nu}(q) & = & \left[\Pi^{\mathrm{T}}_{\mu\nu}(q)+
	\lambda_{\mathrm{St}}
	\Pi^{\mathrm{L}}_{\mu\nu}(q)\right](k^{2}-\vec{q}^{\,2})\Theta(k^{2}-\vec{q}^{\,2})\ ,
	\label{eqn:RV}
\end{IEEEeqnarray}
where we have defined the transversal and longitudinal projection operators $\Pi^{\mathrm{T}}
_{\mu\nu}(q)=\delta_{\mu\nu}-q_{\mu}q_{\nu}/q^{2}$ 
and $\Pi^{\mathrm{L}}_{\mu\nu}(q)=q_{\mu}q_{\nu}/q^{2}$, respectively. Since $\partial_{k}
\Gamma_{k}=\mathcal{V}\partial_{k}U_{k}$ for spatially 
uniform field configurations $\Phi_{i}(x)=\Phi_{i}$, and since in the limit $V\rightarrow 
\infty$ we have $1/V\sum_{\vec{q}}\rightarrow 
\int{\mathrm{d}^{3}q/(2\pi)^{3}}$, using Eq.~(\ref{eqn:RS}) as well as Eq.~(\ref{eqn:RV}), 
Eq.~(\ref{eqn:floweq}) becomes:
\begin{equation}
	\partial_{k}U_{k}=T\sum_{n}\int_{V(k)}\frac{\mathrm{d}^{3}q}{(2\pi)^{3}}k\left\lbrace\tr
	\bar{\textbf{G}}^{\mathrm{S}}_{k} 
	 (\omega_{n},\vec{q})+\left[\Pi^{\mathrm{T}}_{\mu\nu}(q)+\lambda_{\mathrm{St}}
	\Pi^{\mathrm{L}}_{\mu\nu}(q)\right] 
	\tr \bar{\textbf{G}}^{\mathrm{V}}_{k,\nu\mu} (\omega_{n},\vec{q})\right\rbrace
	\ ,\label{eqn:flowU}
\end{equation}
where $\bar{\mathbf{G}}=\mathcal{V}\mathbf{G}$ and $V(k)$ denotes the spherical volume with 
radius $k$ in three-momentum space.

\section{Results}\label{sec:results}

In this section, we numerically solve the FRG flow equations in LPA for the eLSM introduced in 
Sec.\ \ref{sec:LSM} for three scenarios: (i) the $U(2)_{R}\times U(2)_{L}$-symmetric case, 
(ii) the case with $U(1)_{A}$ anomaly, and (iii) the case with $U(1)_{A}$ anomaly and ESB. 
To this end, we discuss the expansion of the potential $U_{k}$ in terms of the respective invariants 
under the given symmetry and fix the bare couplings in the UV such that the 
renormalization flow produces reasonable values for the physical observables in the IR. 
From the behavior of the chiral order parameter $\braket{\sigma}_0$ as a function of
temperature we infer the order of the phase transition and illustrate the restoration of 
chiral symmetry by computing various mesonic screening masses.

Let us remark that there are, in principle, two different strategies to proceed \cite{Schaefer2005}. In 
the first strategy one expands the potential $U_{k}$ around a (local) minimum. 
The advantage of this method is that one has to solve only a few flow equations (one for 
each coupling and an additional one for the scale-dependent order parameter). In 
doing so, however, we can only deduce the potential 
right at a local minimum. It is not clear whether this local minimum is also
the global one. Especially in the case of a first-order 
transition with two emerging minima, it is crucial not just to know the potential at the 
expansion point but also at any other local extremum. 

To overcome this difficulty, in this paper we follow the second strategy, where $U_{k}$ is discretized 
on a grid. Here we gain information about the
entire form of the potential, but this strategy needs a lot of computational power as we have to solve 
flow equations for each grid point. Nevertheless, 
we utilize this approach because it allows us to figure out the transition order in a 
comparatively quick and uncomplicated fashion. We tune the 
potential in such a way that the physical configuration is located at $\braket{\sigma}_{0}
= \sqrt{Z_{\pi}}\, f_{\pi}$ for $k\rightarrow 0$. 

Remember also that the effective action $\Gamma_{k}$ is a functional of the classical fields 
$\phi_{k,i}=\braket{\varphi_{i}}$ and 
$\mathcal{A}_{k,i,\mu}=\braket{A_{i,\mu}}$, cf.\ Eq.~(\ref{eqn:Legendre}), but for the
sake of simplicity the brackets indicating expectation values will be omitted in the 
following, e.g.\ $\braket{\eta} \rightarrow \eta$. Another point is that we are setting certain  
fields to zero after the calculation of $\mathbf{\Gamma}_{k}^{(2)}$, since we only
need to consider as many fields to be non-vanishing as there are independent invariants. 

\subsection{Chiral limit without anomaly}\label{sec:u2u2}

For zero quark masses and in the absence of $U(1)_{A}$-symmetry breaking 
($h_{0}^{0}=0$ as well as $c_{A}=0$), the Lagrangian 
(\ref{eqn:Lglobal}) is invariant under the full $U(2)_R \times U(2)_L$ symmetry. 
In the LPA neither wave-function renormalization nor 
momentum-dependent interactions are taken into account. 
Since the involved (axial\hbox{-})\-vector mesons typically 
have a mass of $\lesssim 1.2$ GeV, an ultraviolet 
cutoff of $\Lambda=1.2$ GeV is chosen. After substituting the fields by their expectation 
values, in compliance with Eq.~(\ref{eqn:derexpLPA}) 
we derive from Eq.~(\ref{eqn:Lglobal}) the following effective average action at the UV 
scale:
\begin{IEEEeqnarray}{rCl}
	\Gamma_{\Lambda} & = & \int_{x}\Bigg\lbrace\frac{1}{2}\partial_{\mu}\sigma\partial_{\mu}
	\sigma
	+\frac{1}{2}\partial_{\mu}\vec{a}_{0}\cdot\partial_{\mu}\vec{a}_{0}
	+\frac{1}{2}\partial_{\mu}\eta\partial_{\mu}\eta+\frac{1}{2}\partial_{\mu}\vec{\pi}\cdot
	\partial_{\mu}\vec{\pi}
	+\frac{1}{4}\left(\partial_{\mu}\omega_{\nu}-\partial_{\nu}\omega_{\mu}\right)^{2}
	+\frac{1}{4}\left(\partial_{\mu}\vec{\rho}_{\nu}-\partial_{\nu}\vec{\rho}_{\mu}
	\right)^{2}\nonumber\\
	& & \quad \; +\frac{1}{4}\left(\partial_{\mu}f_{1\nu}-\partial_{\nu}f_{1\mu}\right)^{2}
	+\frac{1}{4}\left(\partial_{\mu}\vec{a}_{1\nu}-\partial_{\nu}\vec{a}_{1\mu}\right)^{2}
	+\frac{\lambda_{\mathrm{St}}}{2}\left[\left(\partial_{\mu}\omega_{\mu}\right)^{2}+
	\left(\partial_{\mu}
	\vec{\rho}_{\mu}\right)^{2}
	+\left(\partial_{\mu}f_{1\mu}\right)^{2}+\left(\partial_{\mu}\vec{a}_{1\mu}\right)^{2}
	\right]\nonumber\\
	& & \quad \;  +\, c_{1,\Lambda}\xi_{1}+c_{2,\Lambda}\xi_{1}^{2}+c_{3,\Lambda}\xi_{2}+c_{4,\Lambda}\xi_{3}
	+c_{5,\Lambda}\xi_{4}\Bigg\rbrace\ .\label{eqn:LLPA}
\end{IEEEeqnarray}
In Eq.~(\ref{eqn:LLPA}) we have introduced the $O(8)$ mass invariants $\xi_{1}$ and $\xi_{4}$ 
as well as the other $U(2)_{R}\times U(2)_{L}$-
symmetric expressions $\xi_{2}$, $\xi_{3}$ contained in Eq.~(\ref{eqn:Lglobal}). They are 
linear combinations of the different interaction terms 
in the effective potential, namely:
\begin{IEEEeqnarray}{rCl}
	\xi_{1} & = & \sigma^{2}+\vec{a}_{0}^{2}+\eta^{2}+\vec{\pi}^{2}\ ,\label{eqn:xi1}\\
	\xi_{2} & = & \left(\sigma^{2}+\vec{\pi}^{2}\right)\left(\eta^{2}+\vec{a}_{0}^{2}\right)
	-\left(\sigma\eta-\vec{\pi}\cdot\vec{a}_{0}\right)^{2}\ ,\\
	\xi_{3} & = & \left(\vec{\pi}\cdot\vec{a}_{1\mu}+\eta f_{1\mu}\right)^{2}+\left(\vec{a}
	_{0}\cdot\vec{a}_{1\mu}
	+\sigma f_{1\mu}\right)^{2}\nonumber\\
	& & + \left(\vec{\rho}_{\mu}\times\vec{a}_{0}+\eta\vec{a}_{1\mu}+\vec{\pi}f_{1\mu}
	\right)^{2}
	+\left(\vec{\pi}\times\vec{\rho}_{\mu}+\sigma\vec{a}_{1\mu}+\vec{a}_{0}f_{1\mu}
	\right)^{2}\ ,\\
	\xi_{4} & = & f_{1\mu}^{2}+\vec{a}_{1\mu}^{2}+\omega_{\mu}^{2}+\vec{\rho}_{\mu}^{2}\ .
	\label{eqn:xi4}
\end{IEEEeqnarray}
The scale-dependent couplings are defined as:
\begin{equation}
	c_{1,k}=\frac{m_{0,k}^{2}}{2}\ ,\quad c_{2,k}=\frac{1}{4}\left(\lambda_{1,k}+\frac{1}{2}
	\lambda_{2,k}\right)\ ,
	\quad c_{3,k}=\frac{\lambda_{2,k}}{2}\ ,\quad c_{4,k}=\frac{g_{1,k}^{2}}{2}\ ,\quad 
	c_{5,k}=\frac{m_{1,k}^{2}}{2}\ .
\end{equation}
For the expansion of $U_{k}$ we have to replace all non-vanishing field variables by 
appropriate expressions of the invariants $\lbrace\xi_{1},
\xi_{2},\xi_{3}, \xi_{4}\rbrace$. In order to do 
this, we keep the fields $\sigma$, $a_{0}^{1}$, 
$a_{10}^{1}$, and $\rho_{0}^{1}$ nonzero. All others are set to zero. Solving the four 
Eqs.\ (\ref{eqn:xi1}) -- (\ref{eqn:xi4}) for these four variables yields:
\begin{IEEEeqnarray}{rClCrCl}
	\sigma^{2} & = & \frac{1}{2}\left(\xi_{1}+\sqrt{\xi_{1}^{2}-4\xi_{2}}\right)\ ,
	& \quad & \left(a_{0}^{1}\right)^{2} & = &
	\frac{1}{2}\left(\xi_{1}-\sqrt{\xi_{1}^{2}-4\xi_{2}}\right)\ ,\nonumber\\
	\left(a_{10}^{1}\right)^{2} & = & \frac{\xi_{3}}{\xi_{1}}\ , & \quad &
	\left(\rho_{0}^{1}\right)^{2} & = & \xi_{4}-\frac{\xi_{3}}{\xi_{1}}\ .
\end{IEEEeqnarray}
At first glance, this mapping seems to be singular for $\xi_{1}\rightarrow 0$, but we have 
checked that all parts of the flow equations 
$\propto\xi_{1}^{-1}$ cancel for $\xi_{1}=0$, cf.\ also Refs.\ \cite{Fukushima2011b,Patkos2012}. 
Furthermore, the mapping does not preserve 
Euclidean invariance, since we keep only the $\mu = 0$-component 
of the vector fields $\rho^{1}_{\mu}$ and $a_{1\mu}^{1}$. This 
gives rise to unequal screening masses of some 
vector components with differences $\propto \xi_{2}$, $ \xi_{3}$, or $\xi_{4}$, but this is 
also not relevant since we assume that only the 
sigma field acquires a non-vanishing vacuum expectation value (and hence $\xi_{1}\rightarrow 
\xi_{10}\equiv\sigma_{0}^{2}$ and $\xi_{2}$, 
$\xi_{3}$, $\xi_{4}\rightarrow 0$). This means that we are concerned with a 
one-dimensional investigation along the $\xi_{1}$-axis and 
that, in this limit, Euclidean symmetry is restored again.

Without axial anomaly and quark masses, we expect the chiral phase transition to be of first 
order as argued in Ref.~\cite{Pisarski1984} (among 
many other studies) and summarized by the ``Columbia plot'' \cite{Brown1990,Philipsen2008}. 
Since $\xi_{2}$, $\xi_{3}$, and $\xi_{4}$ are set to 
zero in the end, it is reasonable to truncate $U_{k}$ at linear order in these invariants 
(although the flow equation generates terms of arbitrary order in these invariants), with
coefficients that are functions of $\xi_1$:
\begin{equation}
	U_{k}(\xi_{1},\xi_{2},\xi_{3},\xi_{4})=V_{k}(\xi_{1})+W_{k}(\xi_{1})\xi_{2}+X_{k}
	(\xi_{1})\xi_{3}+Y_{k}(\xi_{1})\xi_{4}\ .
\end{equation}

The physical vacuum is specified by the condition $\partial U_{k}/\partial\sigma=0$ for $
\sigma=\sigma_{0}$, and the squared mass of the $\sigma$ 
field is identical to the curvature of the effective potential:
\begin{IEEEeqnarray}{rCl}
	\left.\frac{\partial U_{k}}{\partial\sigma}\right|_{\sigma=\sigma_{0}} & = & 2\sqrt{\xi_{10}}
	V_{k}'(\xi_{10})=0\ ,\label{eqn:min}\\
	\left.\frac{\partial^{2}U_{k}}{\partial\sigma^{2}}\right|_{\sigma=\sigma_{0}} & = & 
	2V_{k}'(\xi_{10})+4\xi_{10}V_{k}''(\xi_{10})\ .
\end{IEEEeqnarray}
From Eq.\ (\ref{eqn:min}) one sees that, for $\xi_{10} \neq 0$, the minimum of
the effective potential can also be determined from the condition $V_{k}'=0$. 
Once the system changes to the restored phase ($\sigma_{0}^2=\xi_{10} = 0$), $V_{k}'$ can be nonzero, though. 

The explicit flow equations for $V_{k}$, $W_{k}$, 
$X_{k}$, and $Y_{k}$ are obtained by differentiating Eq.~(\ref{eqn:flowU}) and evaluating it 
for the physical configuration:
\begin{IEEEeqnarray}{rCl}
	\partial_{k}V_{k}(\xi_{1}) & = & \left.\partial_{k}U_{k}(\xi_{1},\xi_{2},\xi_{3},\xi_{4})
	\right|_{\xi_{2},\xi_{3},\xi_{4}=0}
	\equiv\frac{1}{2}\left.\tr\left(\mathbf{G}_{k}\partial_{k}\mathbf{R}_{k}\right)\right|
	_{\xi_{2},\xi_{3},\xi_{4}=0}\ ,\\
	\partial_{k}W_{k}(\xi_{1}) & = & \left.\partial_{k}\partial_{\xi_{2}}U_{k}(\xi_{1},
	\xi_{2},\xi_{3},\xi_{4})\right|_{\xi_{2},
	\xi_{3},\xi_{4}=0}
	\equiv\frac{1}{2}\mathcal{V}\left.\tr\left(-\mathbf{G}_{k}\mathbf{\Gamma}^{(3)}_{k,
	\xi_{2}}\mathbf{G}_{k}\partial_{k}
	\mathbf{R}_{k}\right)\right|_{\xi_{2},\xi_{3},\xi_{4}=0}\ ,\\
	\partial_{k}X_{k}(\xi_{1}) & = & \left.\partial_{k}\partial_{\xi_{3}}U_{k}(\xi_{1},
	\xi_{2},\xi_{3},\xi_{4})\right|_{\xi_{2},
	\xi_{3},\xi_{4}=0}
	\equiv\frac{1}{2}\mathcal{V}\left.\tr\left(-\mathbf{G}_{k}\mathbf{\Gamma}^{(3)}_{k,
	\xi_{3}}\mathbf{G}_{k}\partial_{k}
	\mathbf{R}_{k}\right)\right|_{\xi_{2},\xi_{3},\xi_{4}=0}\ ,\\
	\partial_{k}Y_{k}(\xi_{1}) & = & \left.\partial_{k}\partial_{\xi_{4}}U_{k}(\xi_{1},
	\xi_{2},\xi_{3},\xi_{4})\right|_{\xi_{2},
	\xi_{3},\xi_{4}=0}
	\equiv\frac{1}{2}\mathcal{V}\left.\tr\left(-\mathbf{G}_{k}\mathbf{\Gamma}^{(3)}_{k,
	\xi_{4}}\mathbf{G}_{k}\partial_{k}
	\mathbf{R}_{k}\right)\right|_{\xi_{2},\xi_{3},\xi_{4}=0}\ .
\end{IEEEeqnarray}
The equation for $V_{k}$ turns out to be equivalent to the flow equation for free 
fields, as interactions are no longer present for $\xi_{2},\xi_{3},\xi_{4}=0$:
\begin{IEEEeqnarray}{rCl}
  	\partial_{k}V_{k}(\xi_{1}) & = & \frac{Tk^{4}}{6\pi^{2}}\sum_{n}\bigg[\frac{4}
	{\omega_{n}^{2}+k^{2}+2V_{k}'}
	+\frac{1}{\omega_{n}^{2}+k^{2}+2V_{k}'+4\xi_{1} V_{k}''}+\frac{3}{\omega_{n}^{2}
	+k^{2}+2V_{k}'+2\xi_{1} W_{k}}\nonumber\\
	& & \hspace*{1.2cm} +\, \frac{12}{\omega_{n}^{2}+k^{2}+2Y_{k}}+\frac{12}{\omega_{n}^{2}+k^{2}+2(Y_{k}+
	\xi_{1} X_{k})}\nonumber\\
	& & \hspace*{1.2cm}+\, \frac{4}{\omega_{n}^{2}+k^{2}+2Y_{k}/\lambda_{\mathrm{St}}}+\frac{4}{\omega_{n}^{2}
	+k^{2}+2(Y_{k}+\xi_{1}X_{k})/\lambda_{\mathrm{St}}}\bigg]\ .\label{eqn:flowV}
\end{IEEEeqnarray}
Here, explicit $\xi_{1}$-dependences on the right-hand side are implied. The masses
of the fields $\rho$ and $\omega$ 
as well as $f_{1}$ and $a_{1}$ are equal in 
the presence of $SU(2)_{V}$ invariance. Equation~(\ref{eqn:flowV}) reveals some 
characteristics of the considered model: the $4+1+3=8$ 
(pseudo\hbox{-})\-scalar degrees of freedom are represented by the first line. The first term 
of Eq.~(\ref{eqn:flowV}) corresponds to the mass-degenerate $\eta$ and $\vec{\pi}$. 
The second and the third describe the $\sigma$ and the $\vec{a}_{0}$, respectively. 
As a cross-check, note 
that the mass eigenvalues of these spin-zero fields are equal to the ones quoted in 
Ref.~\cite{Fukushima2011b}. The second line corresponds to
the physical degrees of freedom of the (axial\hbox{-})\-vector mesons 
($(4 \times 3) + (4 \times 3) =12+12$ fields), 
whereas the last line corresponds to their unphysical degrees of freedom 
($4  + 4$) introduced
via Stueckelberg's Lagrangian. Obviously, these eight additional degrees of freedom 
decouple from the flow for $\lambda_{\mathrm{St}}\rightarrow 0$. The 
Matsubara sum over $n$ can be carried out analytically, e.g.\ by a contour integral in the 
complex plane \cite{Nieto1995}. The flow of $V_{k}(\xi_{1})$ entangles with the flow 
of $W_{k}(\xi_{1})$, $X_{k}(\xi_{1})$, and $Y_{k}(\xi_{1})$. Thus the flow equations 
for these coefficients 
are necessary to obtain a closed set of differential equations. For the sake of clarity, they 
are presented in Appendix \ref{sec:floweqns}.

Figure~\ref{fig:full}~\textbf{A} summarizes the dependence of $\sigma_{0}$ on $T$. 
The order parameter $\sigma_{0}$ becomes successively smaller and drops 
discontinuously to zero 
at $T_c \simeq 147.4$ MeV, indicating a first-order phase transition. The discontinuity 
occurring at this temperature is marked with a 
dashed line. Figure~\ref{fig:full}~\textbf{B} demonstrates how, 
as the temperature increases, the mesonic screening masses 
of chiral partners approach each other.
These are (i) $\rho$ and $a_{1}$, (ii) $\omega$ and $f_1$ (their masses are not shown explicitly,
since $m_\omega = m_\rho$ and $m_{f_1} = m_{a_1}$), (iii) $\sigma$ and $\pi$, as well as 
(iv) $a_{0}$ and $\eta$ ($m_\eta = m_\pi$, thus we do not show $m_\eta$ explicitly). 
The masses of chiral partners become degenerate at the transition temperature and above. 
Note that solid lines correspond to data interpolated using cubic 
splines. In cases where the data points are explicitly given in terms of colored crosses, 
however, the lines represent a cubic smoothing spline fit. Details are provided in Appendix 
\ref{sec:interpolation}. 

The $\rho$/$a_{1}$ mass increases/decreases before the transition 
point is reached. Pions and $\eta$ are the Goldstone bosons of chiral symmetry breaking,
and thus necessarily massless in the broken phase. This is explained by Eq.~(\ref{eqn:min}): 
if $\xi_{10}\neq 0$, the physical minimum is located at 
the point where $V'_{k}(\xi_{1})=0$. Inspecting Eq.~(\ref{eqn:flowV}),
we see that $V_k'$ is indeed proportional to the (squared) mass of pion and $\eta$.
However,  for $T>T_{c}$ the ground state is characterized by $\xi_{10}=0$, and
thus $V'_{k}(0)$ (i.e., the masses of pions and $\eta$) may differ from zero. 

We tuned the UV-parameters in the vacuum to 
achieve the most ``realistic'' mesonic screening masses and a nonzero value for $\sigma_{0}$ of 
around 147.9 MeV (the tree-level value is $\sigma_0 = \sqrt{Z_\pi}\, f_\pi \simeq 148.8$ 
MeV). Apparently, the IR vacuum masses of the $\sigma$ and $a_{0}$ mesons are far too small 
compared to what we expect from the PDG \cite{Olive2014}, no matter whether we
choose the assignment $\{f_0(500),a_0(980)\}$ or $\{f_0(1370),a_0(1450)\}$. We
will return to a discussion of this issue below.
Furthermore, the masses of $\omega$ and $\rho$ are too heavy. 
The ratio between the masses of $\rho$ and $a_{1}$ is smaller than expected 
(experimentally it should be around 1.6).
\begin{figure}[ht]
	\centering
		\includegraphics{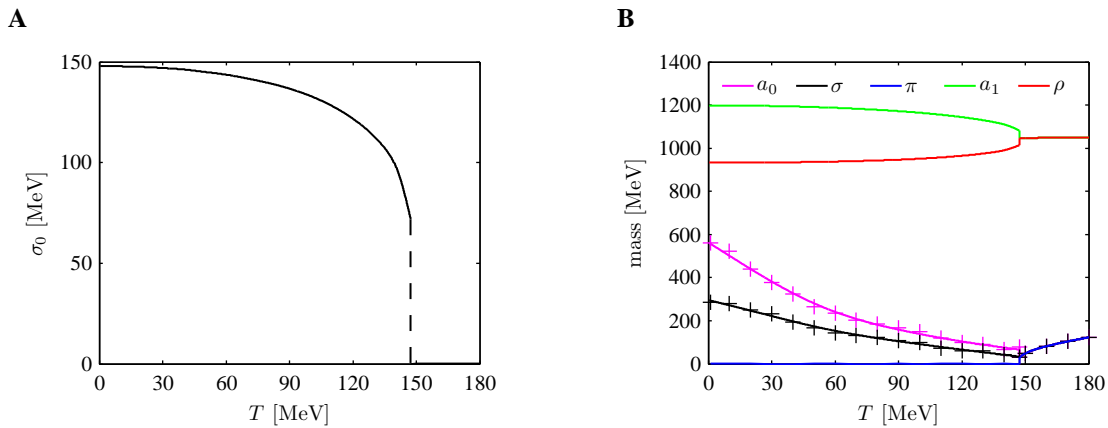}
	\caption{Phase transition in the eLSM with $U(2)_{R}\times U(2)_{L}$ symmetry. \textbf{A} 
	Order parameter $\sigma_{0}$ as a function of temperature. 
	\textbf{B} Screening masses as a function of temperature.}
	\label{fig:full}
\end{figure}

\subsection{Chiral limit with $U(1)_A$ anomaly}\label{sec:anomaly}
As a second scenario we want to study how the $U(1)_{A}$ anomaly influences 
the order of the transition and the mesonic masses. The 
coupling $c_{A}$ in Eq.~(\ref{eqn:Lglobal}) is now nonzero and quantifies the strength of the 
$U(1)_{A}$-symmetry breaking. Proceeding similarly 
as above, $\Gamma_{\Lambda}$ is slightly modified:
\begin{IEEEeqnarray}{rCl}
	\Gamma_{\Lambda} & = & \int_{x}\Bigg\lbrace\frac{1}{2}\partial_{\mu}\sigma\partial_{\mu}
	\sigma+\frac{1}{2}
	\partial_{\mu}\vec{a}_{0}\cdot\partial_{\mu}\vec{a}_{0}
	+\frac{1}{2}\partial_{\mu}\eta\partial_{\mu}\eta+\frac{1}{2}\partial_{\mu}\vec{\pi}\cdot
	\partial_{\mu}\vec{\pi}
	+\frac{1}{4}\left(\partial_{\mu}\omega_{\nu}-\partial_{\nu}\omega_{\mu}\right)^{2}
	+\frac{1}{4}\left(\partial_{\mu}\vec{\rho}_{\nu}-\partial_{\nu}\vec{\rho}_{\mu}
	\right)^{2}\nonumber\\
	& & \quad \; +\frac{1}{4}\left(\partial_{\mu}f_{1\nu}-\partial_{\nu}f_{1\mu}\right)^{2}
	+\frac{1}{4}\left(\partial_{\mu}\vec{a}_{1\nu}-\partial_{\nu}\vec{a}_{1\mu}\right)^{2}
	+\frac{\lambda_{\mathrm{St}}}{2}\left[\left(\partial_{\mu}\omega_{\mu}\right)^{2}
	+\left(\partial_{\mu}\vec{\rho}_{\mu}\right)^{2}+\left(\partial_{\mu}f_{1\mu}\right)^{2}+
	\left(\partial_{\mu}
	\vec{a}_{1\mu}\right)^{2}\right]\nonumber\\
	& & \quad \; +\, \bar{c}_{1,\Lambda}\bar{\xi}_{1}+\bar{c}_{2,\Lambda}\bar{\xi}_{1}^{2}+\bar{c}_{3,
	\Lambda}\bar{\xi}_{1}\bar{\xi}_{2}
	+\bar{c}_{4,\Lambda}\bar{\xi_{2}}+\bar{c}_{5,\Lambda}\bar{\xi}_{2}^{2}
	+\bar{c}_{6,\Lambda}\bar{\xi_{3}}+c_{4,\Lambda}\xi_{3}+c_{5,\Lambda}\xi_{4}
	\Bigg\rbrace\ ,\label{eqn:Lanomaly}
\end{IEEEeqnarray}
with the new invariants $\bar{\xi}_{1}$, $\bar{\xi}_{2}$, $\bar{\xi}_{3}$:
\begin{IEEEeqnarray}{rCl}
	\bar{\xi}_{1} & = & \sigma^{2}+\vec{\pi}^{2}\ ,\\
	\bar{\xi}_{2} & = & \eta^{2}+\vec{a}_{0}^{2}\ ,\\
	\bar{\xi}_{3} & = & \left(\sigma\eta-\vec{\pi}\cdot\vec{a}_{0}\right)^{2}\ .
\end{IEEEeqnarray}
The former invariants $\xi_{1}$ and $\xi_{2}$ are functions of the new invariants
$\bar{\xi}_{1}$, $\bar{\xi}_{2}$, and $\bar{\xi}_{3}$:
$\xi_1 = \bar{\xi}_1 + \bar{\xi}_2$, $\xi_2 = \bar{\xi}_1 \bar{\xi}_2 - \bar{\xi}_3$. 
The origin of the new invariants is the $U(1)_A$-symmetry breaking term
$\sim \det\, \Sigma + \det \, \Sigma^\dagger$ in Eq.~(\ref{eqn:Lglobal}).
The invariants $\xi_{3}$ and $\xi_{4}$ remain unchanged. 
The scale-dependent couplings of the effective potential are 
now defined as follows:
\begin{IEEEeqnarray}{rClCrClCrCl}
	\bar{c}_{1,k} & = & \frac{1}{2}\left(m_{0,k}^{2}-c_{A,k}\right)\ , & \quad & 
	\bar{c}_{2,k} & = & \frac{1}{4}\left(\lambda_{1,k}+\frac{1}{2}\lambda_{2,k}\right)\ , & 
	\quad & 
	\bar{c}_{3,k} & = & \frac{1}{2}\left(\lambda_{1,k}+\frac{3}{2}\lambda_{2,k}\right)\ ,
	\nonumber\\
	\bar{c}_{4,k} & = & \frac{1}{2}\left(m_{0,k}^{2}+c_{A,k}\right)\ , & \quad &
	\bar{c}_{5,k} & = & \frac{1}{4}\left(\lambda_{1,k}+\frac{1}{2}\lambda_{2,k}\right)\ , & 
	\quad & 
	\bar{c}_{6,k} & = & -\frac{\lambda_{2,k}}{2}\ ,\nonumber\\
	c_{4,k} & = & \frac{g_{1,k}^{2}}{2}\ , & \quad & 
	c_{5,k} & = & \frac{m_{1,k}^{2}}{2}\ .
\end{IEEEeqnarray}
We again want to expand $U_{k}$ in terms of $\bar{\xi}_{1},\ldots,\xi_{4}$. For five 
different invariants we have to keep at least five fields 
nonzero. In addition to $\sigma$, $a_{0}^{1}$, $a_{10}^{1}$, and $\rho_{0}^{1}$ we decided 
to take the $\eta$ field into account and map those 
variables onto the set $\lbrace\bar{\xi}_{1}, \bar{\xi}_{2},\bar{\xi}_{3},\xi_{3},\xi_{4}\rbrace$:
\begin{equation}
	\sigma^{2} = \bar{\xi}_{1}\ ,\quad \left(a_{0}^{1}\right)^{2} =
	\bar{\xi}_{2}
	-\frac{\bar{\xi}_{3}}{\bar{\xi}_{1}}\ ,
	\quad \eta^{2}= \frac{\bar{\xi}_{3}}{\bar{\xi}_{1}}\ ,\quad \left(a_{10}^{1}
	\right)^{2}= 
	\frac{\xi_{3}}{\bar{\xi}_{1}+\bar{\xi}_{2}}\ ,
	\quad \left(\rho_{0}^{1}\right)^{2}= \xi_{4}-\frac{\xi_{3}}{\bar{\xi}_{1}+
	\bar{\xi}_{2}}\ .
\end{equation}
Similarly to Sec.\ \ref{sec:u2u2}, the potential is now discretized with respect to
$\bar{\xi}_{1}$ and 
one assumes that the physical minimum is attained 
for $\bar{\xi}_{10}\equiv\sigma_{0}^{2}$ and $\bar{\xi}_{2},
\bar{\xi}_{3}= 0$. Thus, in the ansatz for the 
effective potential, we restrict ourselves to the linear order in $\bar{\xi}_{2}$, 
$\bar{\xi}_{3}$, $\xi_{3}$, and $\xi_{4}$:
\begin{equation}
	U_{k}(\bar{\xi}_{1},\bar{\xi}_{2},\bar{\xi}_{3},\xi_{3},\xi_{4})=\bar{V}_{k}(\bar{\xi}
	_{1})+\bar{W}_{k}(\bar{\xi}_{1})
	\bar{\xi}_{2}+\bar{X}_{k}(\bar{\xi}_{1})\bar{\xi}_{3}+\bar{Y}_{k}(\bar{\xi}_{1})\xi_{3}
	+\bar{Z}_{k}(\bar{\xi}_{1})\xi_{4}
	\ .\label{eqn:truncanoma}
\end{equation}
The flow equation for 
$\bar{V}_{k}(\bar{\xi}_{1})$ reads:
\begin{IEEEeqnarray}{rCl}
  	\partial_{k}\bar{V}_{k}(\bar{\xi}_{1}) & = & \frac{Tk^{4}}{6\pi^{2}}\sum_{n}
  	\bigg[\frac{3}{\omega_{n}^{2}+k^{2}+2\bar{V}_{k}'}
	+\frac{1}{\omega_{n}^{2}+k^{2}+2\bar{V}_{k}'+4\bar{\xi}_{1} \bar{V}_{k}''}+\frac{3}
	{\omega_{n}^{2}+k^{2}+2\bar{W}_{k}}
	\nonumber\\
	& & +\frac{1}{\omega_{n}^{2}+k^{2}+2(\bar{W}_{k}+\bar{\xi}_{1}\bar{X}_{k})}+\frac{12}
	{\omega_{n}^{2}+k^{2}+2\bar{Z}_{k}}+\frac{12}{\omega_{
	n}^{2}+k^{2}+2(\bar{Z}_{k}+\bar{\xi}_{1}\bar{Y}_{k})}\nonumber\\
	& & +\frac{4}{\omega_{n}^{2}+k^{2}+2\bar{Z}_{k}/\lambda_{\mathrm{St}}}+\frac{4}{\omega_{n}^{2}
	+k^{2}+2(\bar{Z}_{k}
	+\bar{\xi}_{1}\bar{Y}_{k})/\lambda_{\mathrm{St}}}\bigg]\ .\label{eqn:flowVanomaly}
\end{IEEEeqnarray}
Note that the $\eta$ meson now attains a mass different
from that of the pions: it becomes massive in the 
spontaneously broken phase, as it should be, since $U(1)_A$ is explicitly broken and there
are only three (and no longer four) Goldstone bosons.

\begin{figure}[ht]
	\centering
		\includegraphics{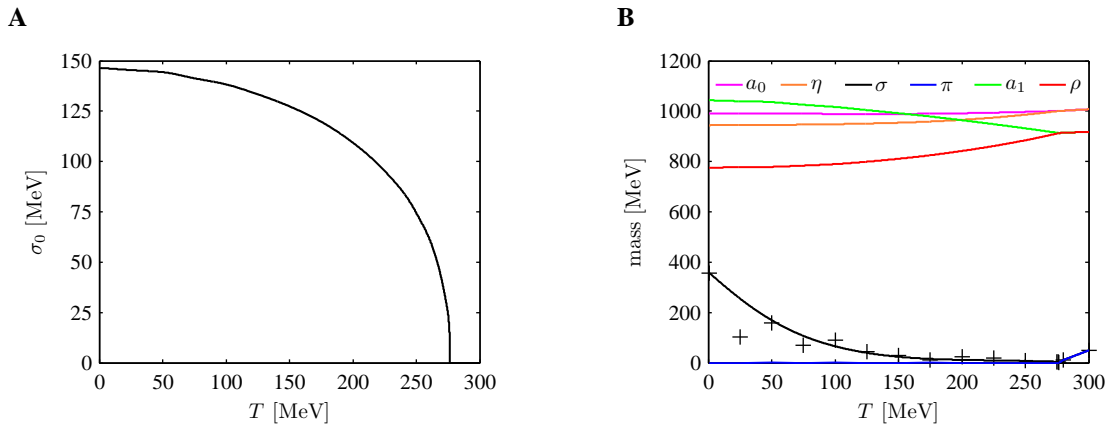}
	\caption{Phase transition in the eLSM with $SU(2)_{V}\times SU(2)_{A}\times 
	U(1)_{V}$ symmetry. \textbf{A} Order
	parameter $\sigma_{0}$ as a function of the temperature $T$.
	\textbf{B} Screening masses as a function of temperature.}
	\label{fig:anomaly}
\end{figure}

Again we tune the bare parameters such that most realistic meson masses are obtained in the 
IR. The order parameter $\sigma_{0}$ continuously decreases with 
increasing $T$, until it vanishes at a critical temperature of approximately 276.5 MeV, see 
Fig.~\ref{fig:anomaly}~\textbf{A}. 
We conclude that the axial anomaly turns the first-order transition, found in the 
$U(2)_R \times U(2)_L$-symmetric theory, into a second-order 
phase transition with a significantly higher critical temperature. Figure~
\ref{fig:anomaly}~\textbf{B} shows the evolution of the masses of
(pseudo\hbox{-})\-scalar and (axial\hbox{-})\-vector mesons. 
Again, there are four different pairs of chiral 
partners, ($\rho$,$a_{1}$), ($\omega$,$f_{1}$) (the masses of which are identical to the
corresponding mesons in the first pair and thus not shown explicitly), ($\sigma$,$\pi$), 
as well as ($\eta$,$a_{0}$). The masses of chiral partners
become degenerate at the transition temperature. 
The pions assume a non-vanishing mass above the critical temperature for the same reason as 
discussed in the previous section.
At zero temperature, the $\rho (\omega)$ meson mass is close to its physical value, 
but the $a_1(f_1)$ mass is too small. 
The mass difference between vector ($\omega$, $\rho$) and axial-vector mesons ($f_{1}$, 
$a_{1}$) comes out to be $\simeq 268.7$ MeV and 
is of the same magnitude as for the $U(2)_{R}\times 
U(2)_{L}$-symmetric case ($\simeq 265.4$ MeV).
The $\eta$-meson mass is too large, but
remember that the $\eta$ in our case consists only of nonstrange quarks, while
the physical $\eta$ is an admixture of nonstrange and strange quarks. 
The vacuum mass of the $\sigma$ of around 357.4 MeV is now only slightly smaller than the
experimental value for the mass of $f_0(500)$, while the vacuum mass of $a_0$ is
very close to its experimental value. The data points for the mass of the $\sigma$ fluctuate
strongly as a function of temperature, since the potential is rather flat in this case and its
curvature (the squared $\sigma$ mass) is rather hard to determine numerically with reasonable
accuracy.

\subsection{Explicitly broken chiral symmetry with anomaly}\label{sec:esb}
Finally, we apply our FRG analysis to the case of ESB due to 
nonzero and degenerate quark masses ($h_{0}^{0}\neq 
0$). The truncation of the effective potential (\ref{eqn:truncanoma}) is still valid. In the 
case of ESB, the root of $\bar{V}'_{k}$ 
no longer coincides with the one of $\partial U_{k}/\partial\sigma$ for nonzero 
vacuum expectation values of the $\sigma$ field. The 
global minimum $\bar{\xi}_{10}$ must now fulfill the following relation:
\begin{equation}
	2\sqrt{\bar{\xi}_{10}}\bar{V}_{k}'(\bar{\xi}_{10})-h_{0}^{0}=0\ ,\label{eqn:minimum}
\end{equation}
i.e., the expansion coefficients $\bar{W}_{k}$, $\bar{X}_{k}$, etc., are evaluated for 
a shifted $\bar{\xi}_{10}$, producing massive 
pseudo-Goldstone bosons $\pi$ (which have a nonzero mass even for $\sigma_{0}\neq 0$). 

\begin{figure}[ht]
	\centering
		\includegraphics{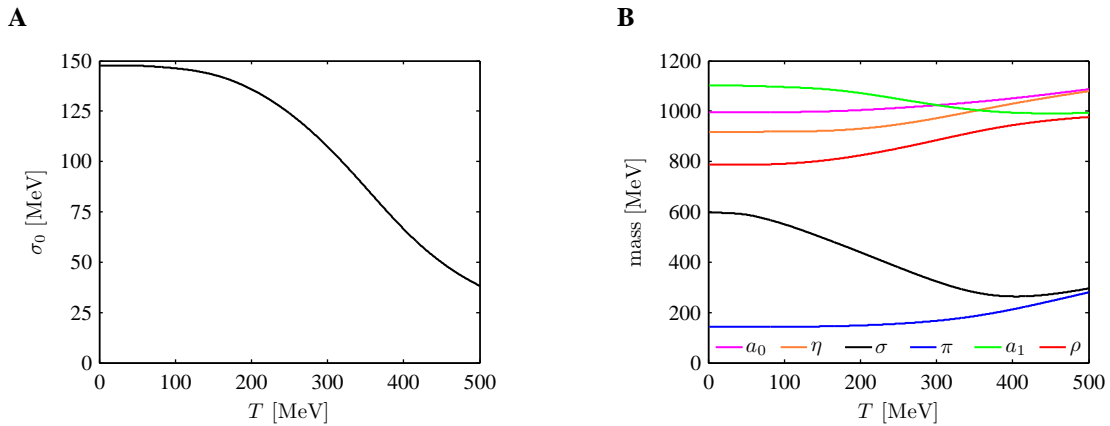}
	\caption{Phase transition in the eLSM with ESB and $U(1)_A$ anomaly. \textbf{A} 
	Order parameter $\sigma_{0}$ as a function of temperature.
	\textbf{B} Screening masses as a function of temperature.}
	\label{fig:esb1}
\end{figure}

According to Eq.~(\ref{eqn:minimum}), the potential never has a global minimum at $\bar{\xi}
_{1}=0$.  Solving the flow 
equations on the $\bar{\xi_{1}}$ grid, the results are 
shown in Fig.~\ref{fig:esb1}. In Fig.~\ref{fig:esb1}~\textbf{A}, we see that $\sigma_{0}$ 
decreases and tends asymptotically towards zero. Consequently, the transition is a crossover
transition. At an 
estimated pseudocritical temperature of $T_{pc}\simeq 354.8$ MeV, the curvature changes its 
sign.
Figure~\ref{fig:esb1}~\textbf{B} reveals that, with the choice $h_{0}^{0}=3\times 10^{6}\ \mathrm{MeV}^{3}$, 
the pions exhibit a nonzero mass of $\simeq 142.5$ MeV in vacuum. The masses of chiral partners 
approach each other, but do not become identical. We 
further observe a dropping $a_{1}$ meson mass but an increasing $\rho$ mass. The vacuum 
$\sigma$ mass of 598.5 MeV is a little larger than the physical value for the $f_{0}
(500)$ resonance. The remaining masses are in good 
agreement with the results of 
Sec.\ \ref{sec:anomaly}. The gap between the masses of the $\rho$ and $a_{1}$ mesons is 
now 314.5 MeV (compared to 268.7 MeV in the previous case).

\section{Summary and Outlook}
\label{sec:discussion}

The QCD transition is commonly associated with the 
restoration of chiral symmetry. Experimentally, this could be detected by
a change of the in-medium masses of (axial-)vector mesons. It is therefore essential to 
include them in a theoretical
analysis. Non-perturbative continuum 
methods, such as the FRG, provide new insights into 
the QCD transition, as they do not rely on weak couplings and 
are applicable at nonzero net-baryon density where lattice QCD suffers
from the fermion-sign problem. 

In this study, we investigated the chiral transition for two flavors by applying the FRG formalism to the eLSM, 
an effective low-energy model for QCD. 
Thus, our work is an extension of many studies involving two-flavor effective models for QCD, see e.g.\
Refs.~\cite{Fukushima2011b,Grahl2013,Fejos2014}, in the sense that vector and 
axial-vector mesonic degrees of freedom are now incorporated into the FRG flow.
In order to derive the FRG flow equations with (axial-)vector mesons, Stueckelberg's Lagrangian has been employed 
\cite{Itzykson1985,Rischke1994,Ruegg2004}. We use the grid method and the LPA to compute the
flow of the effective potential. The order of the phase transition and
the meson screening masses were determined in three different scenarios: 
(i) the chiral limit without $U(1)_A$ anomaly, (ii) the
chiral limit with $U(1)_A$ anomaly, and (iii) the realistic case with nonvanishing quark masses
and $U(1)_A$ anomaly. 

Overall, our numerical results 
are broadly consistent with previous findings.  
Regarding the full $U(2)_{R}\times 
U(2)_{L}$-symmetric theory, cf.\ Sec.~\ref{sec:u2u2}, our conclusion is compatible with Ref.\
\cite{Fejos2014} and the statement of Pisarski and Wilczek 
\cite{Pisarski1984} that the chiral phase transition is of first order for $N_{f}=2$ and 
massless quarks. Reducing the symmetry to 
$SU(2)_{V}\times SU(2)_{A}\times U(1)_{V}$ (Sec.~\ref{sec:anomaly}) turns it into a second-
order transition, which is also in agreement with 
Ref.~\cite{Pisarski1984} as well as with the three-flavor study of Ref.~\cite{Mitter2014}. 
Explicitly breaking chiral symmetry to an exact isospin symmetry 
generates a crossover transition (Sec.~\ref{sec:esb}). 

In comparison with the results from the CJT 
formalism \cite{Strueber2008} or lattice-QCD simulations 
\cite{Fodor2004,Cheng2006,Aoki2006a}, the pseudocritical temperature $T_{pc}\simeq 354.8$ 
MeV of the crossover transition comes out larger, 
cf.\ $T_{pc}\simeq 195$ MeV resp.\ 155 MeV in the aforementioned approaches. It is 
conceivable that this deviation arises from the 
lack of quark fields in our approach, which, being rather light degrees of freedom, evidently contribute 
substantially to the FRG flow \cite{Rennecke2015}, and from the fact that 
presently our approach ignores momentum-dependent vertices and interactions among (axial\hbox{-})\-
vector mesons. It has to be clarified how higher 
orders in the derivative expansion affect the transition. We hope that by additional 
investigations, as stated below, we are able to improve our 
results. 

In all three scenarios studied here, it was demonstrated how the masses of chiral partners become degenerate 
at the phase boundary and beyond, see 
Figs.~\ref{fig:full}~\textbf{B}, \ref{fig:anomaly}~\textbf{B}, and \ref{fig:esb1}~\textbf{B}. 
The mass degeneracy is a necessary condition for 
the restoration of chiral symmetry \cite{Brown1991,Pisarski1995,Harada2006,Friman2011}. 
Let us note that 
in our study the mass of the $a_{1}$ mass decreases towards
the chiral transition, but not the mass of the $\rho$, cf.\ Fig.~\ref{fig:esb1}~\textbf{B}.
In the CJT study 
of Ref.~\cite{Strueber2008} the authors also found an increasing $\rho$ mass towards the 
chiral transition. In principle, 
Ref.~\cite{Pisarski1995} argues that the $\rho$ meson mass has to increase in the framework of a 
gauged two-flavor LSM, but a globally symmetric LSM could 
also allow for a dropping $\rho$ mass.

In the physically most realistic scenario with ESB and $U(1)_A$ anomaly,
the vacuum masses of $(\sigma,a_0)$ come out to be (598.5, 996.3) MeV.
This is in the range of the masses of the light scalar resonances
$\lbrace f_{0}(500),a_{0}(980)\rbrace$. However, 
Refs.~\cite{Parganlija2010,Parganlija2012,Parganlija2013} suggested that
the chiral partners of $\pi$ and $\eta$ should be $\lbrace f_{0}(1370),a_{0}(1450)\rbrace$. 
This seems to be a natural scenario, if the latter are (predominantly) quark-antiquark states.
Then, the light resonances $\lbrace f_{0}(500),a_{0}(980)\rbrace$ are most likely
made of four quarks, e.g.\ in the form of
resonances in the scattering continuum, or even bound states, of two pseudoscalar mesons.
Also bound states of diquark and anti-diquark molecules have been suggested to explain
their nature. As suggested a long time ago by Jaffe \cite{Jaffe}, 
this "tetraquark" interpretation of the light scalar resonances would
naturally explain the "inverse mass ordering" of these states.
By construction, the FRG approach resums correlations of infinite order and thus,
if $\lbrace f_{0}(500),a_{0}(980)\rbrace$ are correlated states of
pseudoscalar mesons, would naturally generate these mesons dynamically. 
This could be an explanation why the masses of $\sigma$ and $a_0$ come
out close to those of $\lbrace f_{0}(500),a_{0}(980)\rbrace$. (In fact, we
were not able to find UV-parameters such that the IR vacuum masses of
$\sigma$ and $a_0$ are close to those of $\lbrace f_{0}(1370),a_{0}(1450)\rbrace$.)
Note that the chiral transition was studied in the presence of both a light and a heavy scalar state
in Ref.\ \cite{Heinz}. There it was shown that there is actually no conflict with the
"tetraquark" scalar state being light and the heavy scalar state being the chiral partner of the pion.

There are many questions left open for future study. E.g.\ one should investigate the 
order of the phase transition as a function of the anomaly 
strength. The first-order transition in Sec.~\ref{sec:u2u2} should smoothly pass into one of 
second order, as shown in Sec.~\ref{sec:anomaly}. 
Moreover, one needs to figure out when exactly the $U(1)_{A}$ anomaly disappears for high 
temperatures. This can be done by assuming
$c_{A}$ to be proportional to an explicitly $T$-dependent instanton density. 
In order to decide whether 
the transition lies in the $O(4)$-universality class 
or not \cite{Pisarski1984,Grahl2013}, the critical exponents have to be calculated. A natural 
next step in our analysis is to account for non-trivial 
wave-function renormalization factors, i.e., going beyond the LPA. 
One can readily extend our investigations to $N_f=3$ quark flavors, but in this case
one has an additional order parameter (the strange condensate) which necessitates
the use of a two-dimensional grid \cite{Mitter2014} and thus considerably
increases the numerical effort.
Baryonic degrees of freedom also play 
an important role for dilepton production 
\cite{Rapp2000}. Therefore, it is mandatory to involve them in the FRG flow. The first 
candidate for such an extension would be the nucleon 
and its chiral partner \cite{DeTar1989,Wilms2007} -- if the latter exists \cite{Glozman2007}. 
As this partner would probably be heavier than the 
$\Delta$ baryons, one should furthermore include spin-3/2 resonances and their chiral partners 
\cite{Jido2000}. Finally, another topic for future studies is to
introduce quarks in the FRG. The reason for this is that, as mentioned above, 
a large contribution to the FRG flow usually comes from the quarks, while the vector and 
axial-vector mesons are suppressed due to their large
mass. 

\section{Acknowledgments}
The authors thank Jan M.\ Pawlowski, Robert D.\ Pisarski, 
Bernd-Jochen Schaefer, Lorenz von Smekal, and Mario 
Mitter for valuable discussions.

\begin{appendix}
\section{Flow equations}\label{sec:floweqns}
The flow equations for the expansion coefficients of the effective potential $U_{k}$ 
are generally of the form:
\begin{equation}
	\partial_{k}F_{k,i}=\frac{T}{2\pi^{2}}\sum_{n}\int_{0}^{k}\mathrm{d}q\ \vec{q}^{\,2}
	f_{k,i}(F_{k},F_{k}',
	F_{k}'',\omega_{n}^{2},\vec{q}^{\,2})\ ,
\end{equation}
with $F_{k}=(V_{k},W_{k},\ldots)$, $f_{k}=(v_{k},w_{k},\ldots)$, and $q=|\vec{q}^{\,}|$. Dependences on $\xi_{1},
\bar{\xi}_{1}$ were omitted. The derivatives $F_{k}'=\partial F_{k}/\partial\xi_{1}$ and $F_{k}''=
\partial^{2} F_{k}/(\partial\xi_{1})^{2}$ (or $F_{k}'=\partial F_{k}/\partial\bar{\xi_{1}}$ 
and $F_{k}''=\partial^{2} F_{k}/(\partial\bar{\xi_{1}})^{2}$, respectively)
are approximated by finite differences, which leads to a solvable system:
\begin{equation} \label{eqn:Flow}
	\partial_{k}F_{k,i}=\frac{T}{2\pi^{2}}\sum_{n}\int_{0}^{k}\mathrm{d}q\ \vec{q}^{\,2}
	f_{k,i}(F_{k},\omega_{n}^{2},
	\vec{q}^{\,2})\ .
\end{equation}
The functions $f_{k,i}$ are specified in subsections \ref{sec:flowwithout} and 
\ref{sec:flowwith}. The sum over Matsubara frequencies is performed analytically, e.g.
\begin{IEEEeqnarray}{rCl}
	\sum_{n}\int_{0}^{k}\mathrm{d}q\ q^{4}\frac{1}{E_{\sigma}^4\left(\omega _n^2
	+q^{2}\right)} & = & 
  	\int_{0}^{k}\mathrm{d}q\ q^{4}\Biggg[
  	\frac{\coth \left(\frac{q}{2 T}\right)}{2 q T \left(k^2+4 \xi _1 V_k''
  	+2 V_k'-q^2\right){}^2}\nonumber\\
  	& & -\frac{\text{csch}^2\left(\frac{\sqrt{k^2+4 \xi _1 V_k''+2 V_k'}}{2 T}\right)}{8 T^2 
  	\left(k^2+4 \xi _1 V_k''+2 V_k'-q^2\right){}^2}\nonumber\\
  	& & +\frac{q^2 \text{csch}^2\left(\frac{\sqrt{k^2+4 \xi _1 V_k''+2 V_k'}}{2 T}\right)}{8 
  	T^2 \left(k^2+4 \xi _1 V_k''+2 V_k'\right) \left(k^2+4 \xi
   	_1 V_k''+2 V_k'-q^2\right){}^2}\nonumber\\
    & & + \frac{q^2 \sinh \left(\frac{\sqrt{k^2+4 \xi _1 V_k''+2 V_k'}}{T}\right) \text{csch}
    ^2\left(\frac{\sqrt{k^2+4 \xi _1 V_k''+2 V_k'}}{2
   	T}\right)}{8 T \left(k^2+4 \xi _1 V_k''+2 V_k'\right){}^{3/2} \left(k^2+4 
   	\xi _1 V_k''+2 V_k'-q^2\right){}^2}\nonumber\\
   	& & -\frac{3 \sinh \left(\frac{\sqrt{k^2+4 \xi _1 V_k''+2 V_k'}}{T}\right) 
   	\text{csch}^2\left(\frac{\sqrt{k^2+4 \xi _1 V_k''+2 V_k'}}{2 T}\right)}{8
   	T \sqrt{k^2+4 \xi _1 V_k''+2 V_k'} \left(k^2+4 \xi _1 V_k''+2 V_k'-q^2\right){}^2}   	
  	\Biggg]\ .
\end{IEEEeqnarray}
Where necessary, the momentum integration is performed by using numerical quadrature.

The potentials $V_{k}$ and $\bar{V}_{k}$ are initialized as follows:
\begin{IEEEeqnarray}{rCl}
	V_{\Lambda}(\xi_{1}) & = & a_{\Lambda}\left(\xi_{1}-b_{\Lambda}^{2}\right)^{2}\ ,\\
	\bar{V}_{\Lambda}(\bar{\xi}_{1}) & = & \bar{a}_{\Lambda}\left(\bar{\xi}_{1}
	-\bar{b}_{\Lambda}^{2}\right)^{2}\ ,
\end{IEEEeqnarray}
where $a_{\Lambda}=4.0$, $b_{\Lambda}=297.6$ MeV, $\bar{a}_{\Lambda}=5.5$, 
and $\bar{b}_{\Lambda}=255.0$ MeV. Furthermore, we have 
$W_{\Lambda}(\xi_{1})=30$, $X_{\Lambda}
(\xi_{1})=20$, $Y_{\Lambda}(\xi_{1})=2.87\times 10^{5}\ \mathrm{MeV}^{2}$, $\bar{W}_{\Lambda}
(\bar{\xi}_{1})=2.5\times 10^{5}\ 
\mathrm{MeV}^{2}$, $\bar{X}_{\Lambda}(\bar{\xi}_{1})=-1.0$, $\bar{Y}_{\Lambda}(\bar{\xi}
_{1})=31.0$, and $\bar{Z}_{\Lambda}=1.27\times 10^{5}\ 
\mathrm{MeV}^{2}$. Under the assumption of ESB, the UV parameters change to: $\bar{W}
_{\Lambda}(\bar{\xi}_{1})=2.0\times 10^{5}\ 
\mathrm{MeV}^{2}$, $\bar{Y}_{\Lambda}(\bar{\xi}_{1})=45.0$, and $\bar{Z}_{\Lambda}=1.1\times 10^{5}\ 
\mathrm{MeV}^{2}$, while the rest remain identical.

\subsection{Flow equations without $U(1)_A$ anomaly}\label{sec:flowwithout}
In the case without $U(1)_A$ anomaly, the quantities $f_{k,i}$ in Eq.\ (\ref{eqn:Flow}) read
\begin{IEEEeqnarray}{rCl}
	v_{k} & = & k \left(\frac{3}{E_{a_0}^2}+\frac{12}{E_{a_1}^2}+\frac{12}{E_{\rho }^2}+
	\frac{1}{E_{\sigma }^2}
	+\frac{4}{E_{\pi }^2}\right)\ ,\\
	w_{k} & = & \frac{1}{2}k\Bigg\lbrace-\frac{4 \left[2 V_k'' \left(8 W_k-3 \xi _1 
	W_k'\right)+15 \xi _1 W_k W_k'+4
	\xi _1^2 W_k'^{2}+W_k^2\right]}{E_{a_0}^4 \xi _1 \left(W_k-2 V_k''\right)}+\frac{24 X_k}
	{E_{a_1}^4 \xi _1}
	+\frac{192 X_k^2}{E_{a_1}^6}\nonumber\\
	& & -\frac{4 \left[2 V_k'' \left(2 \xi _1^{2} W_k''+9\xi _1 W_k'+6 W_k\right)+\xi _1 W_k 
	\left(3
  	W_k'-2 \xi _1 W_k''\right)+4 \xi _1^2 W_k'^{2}+3 W_k^2\right]}{\xi _1 E_{\sigma }^4 
  	\left(2V_k''-W_k\right)}\nonumber\\
  	& & -\frac{8 \left(2 \xi _1 W_k'+W_k\right)}{E_{\pi }^4 \xi _1}+\frac{16 W_k^2}{E_{\pi }
  	^6}-\frac{24
  	X_k}{\xi _1 E_{\rho }^4}\Bigg\rbrace\ ,\\
	x_{k} & = & \frac{1}{2}k\biggg\lbrace-\frac{16 \vec{q}^{\,2} X_k^2}{E_{\pi }^4 
  	E_{a_1}^2 \left(\omega _n^2+\vec{q}^{\,2}\right)}+
	\frac{32\vec{q}^{\,2} X_k^2}{E_{a_0}^4 E_{a_1}^2 \left(\omega _n^2+\vec{q}^{\,2}\right)}-
	\frac{16 \vec{q}^{\,2} X_k^2}
	{E_{\pi }^2 E_{a_1}^4 \left(\omega _n^2+\vec{q}^{\,2}\right)}+\frac{32 \vec{q}^{\,2} 
	X_k^2}{E_{a_0}^2 E_{a_1}^4 \left(\omega
  	_n^2+\vec{q}^{\,2}\right)}\nonumber\\
  	& & +\frac{32 \vec{q}^{\,2} \left(\xi _1 X_k'+X_k+Y_k'\right){}^2}{E_{a_1}^2 
  	E_{\sigma }^4 \left(\omega
  	_n^2+\vec{q}^{\,2}\right)}+\frac{32 \vec{q}^{\,2} \left(\xi _1 X_k'+X_k+Y_k'\right){}^2}
  	{E_{a_1}^4 E_{\sigma }^2 \left(\omega
  	_n^2+\vec{q}^{\,2}\right)}-\frac{16 \vec{q}^{\,2} X_k^2}{E_{a_0}^4 \left(\omega _n^2+
  	\vec{q}^{\,2}\right) \left(\xi _1
  	X_k+Y_k\right)}\nonumber\\
  	& & +\frac{4 \left[X_k \left(\frac{4 X_k}{\xi _1 X_k+Y_k}+\frac{1}{\xi _1}\right)-3
  	X_k'\right]}{E_{a_0}^4}+\frac{16 \vec{q}^{\,2} X_k^2}{E_{\pi }^4 E_{\rho }^2 
  	\left(\omega _n^2+\vec{q}^{\,2}\right)}
	+\frac{16 \vec{q}^{\,2} X_k^2}{E_{\pi }^2 E_{\rho }^4 \left(\omega _n^2+\vec{q}^{\,2}
	\right)}\nonumber\\
  	& & -\frac{16 \vec{q}^{\,2} \left(\xi _1
  	X_k'+X_k+Y_k'\right){}^2}{E_{\sigma }^4 \left(\omega _n^2+\vec{q}^{\,2}\right) 
  	\left(\xi _1 X_k+Y_k\right)}-\frac{8 \xi _1
  	\vec{q}^{\,2} X_k^3}{E_{\pi }^4 Y_k \left(\omega _n^2+\vec{q}^{\,2}\right) 
  	\left(\xi _1 X_k+Y_k\right)}-\frac{32 \vec{q}^{\,2}
  	Y_k'^{2}}{E_{\rho }^4 E_{\sigma }^2 \left(\omega _n^2+\vec{q}^{\,2}\right)}
  	-\frac{32 \vec{q}^{\,2}
  	Y_k'^{2}}{E_{\rho }^2 E_{\sigma }^4 \left(\omega _n^2+\vec{q}^{\,2}\right)}\nonumber\\
  	& & +\frac{16 \vec{q}^{\,2}
  	Y_k'^{2}}{E_{\sigma }^4 Y_k \left(\omega _n^2+\vec{q}^{\,2}\right)}+\frac{4 
  	\left[-2 \xi _1 X_k''-\frac{X_k}{\xi
  	_1}+\frac{4 \left(\xi _1 X_k'+X_k+Y_k'\right){}^2}{\xi _1 X_k+Y_k}-5 X_k'-\frac{4
  	Y_k'^{2}}{Y_k}\right]}{E_{\sigma }^4}\nonumber\\
  	& & +\frac{8 \left(\frac{\xi _1 X_k^3}{\xi _1 X_k Y_k+Y_k^2}-2
  	X_k'\right)}{E_{\pi }^4}\biggg\rbrace\ ,\\
  	y_{k} & = & \frac{1}{2}k\Bigg\lbrace\frac{16 \xi _1 \vec{q}^{\,2} X_k^2}{E_{\pi }^4 
 	E_{a_1}^2 \left(\omega _n^2
	+\vec{q}^{\,2}\right)}+\frac{16  
  	\xi_1 \vec{q}^{\,2} X_k^2}{E_{\pi }^2 E_{a_1}^4 \left(\omega _n^2+\vec{q}^{\,2}
  	\right)}-\frac{4 \left(2 X_k+3
  	Y_k'\right)}{E_{a_0}^4}-\frac{8 \xi _1 \vec{q}^{\,2} X_k^2}{E_{\pi }^4 \left(\omega _n^2+
  	\vec{q}^{\,2}\right) \left(\xi _1
  	X_k+Y_k\right)}\nonumber\\
  	& & +\frac{32 \xi _1 \vec{q}^{\,2} Y_k'^{2}}{E_{\rho }^4 E_{\sigma }^2 \left(\omega
  	_n^2+\vec{q}^{\,2}\right)}+\frac{32 \xi _1 \vec{q}^{\,2} Y_k'^{2}}{E_{\rho }^2 
  	E_{\sigma }^4 \left(\omega
  	_n^2+\vec{q}^{\,2}\right)}-\frac{16 \xi _1 \vec{q}^{\,2} Y_k'^{2}}{E_{\sigma }^4 
  	Y_k \left(\omega
  	_n^2+\vec{q}^{\,2}\right)}+\frac{8 \left(-\frac{X_k Y_k}{\xi _1 X_k+Y_k}-2 Y_k'\right)}
  	{E_{\pi }^4}\nonumber\\
	& & -\frac{4 \left[Y_k\left(2 \xi _1 Y_k''+Y_k'\right)-4 \xi _1 Y_k'^{2}\right]}
  	{E_{\sigma}^4 Y_k}\Bigg\rbrace\ ,
\end{IEEEeqnarray}
with:
\begin{IEEEeqnarray}{rCl}
	E_{\sigma}^{2} & = & k^{2}+\omega_{n}^{2}+2V_{k}'+4\xi_{1}V_{k}''\ ,\\
	E_{\pi}^{2} & = &  k^{2}+\omega_{n}^{2}+2V_{k}'\ ,\\
	E_{a_{0}}^{2} & = & k^{2}+\omega_{n}^{2}+2V_{k}'+2\xi_{1}W_{k}\ ,\\
	E_{\rho}^{2} & = & k^{2}+\omega_{n}^{2}+2Y_{k}\ ,\\
	E_{a_{1}}^{2} & = & k^{2}+\omega_{n}^{2}+2Y_{k}+2\xi_{1}X_{k}\ .
\end{IEEEeqnarray}
The meson masses are given by:
\begin{IEEEeqnarray}{rCl}
	m_{\sigma} & = & \sqrt{2V_{k}'+4\xi_{1}V_{k}''}\ ,\\
	m_{\pi} & = & \sqrt{2V_{k}'}\ ,\\
	m_{a_{0}} & = & \sqrt{2V_{k}'+2\xi_{1}W_{k}}\ ,\\
	m_{\rho} & = & \sqrt{2Y_{k}}\ ,\\
	m_{a_{1}} & = & \sqrt{2Y_{k}+2\xi_{1}X_{k}}\ .
\end{IEEEeqnarray}

\subsection{Flow equations with $U(1)_A$ anomaly}\label{sec:flowwith}
In the case with $U(1)_A$ anomaly, the quantities $f_{k,i}$ in Eq.\ (\ref{eqn:Flow}) read
\begin{IEEEeqnarray}{rCl}
	\bar{v}_{k} & = & \frac{1}{2} k \left[24 \left(\frac{1}{E_{a_1}^2}+\frac{1}{E_{\rho }^2}
	\right)+\frac{6}{E_{a_0}^2}
	+\frac{2}{E_{\eta }^2}+\frac{2}{E_{\sigma}^2}+\frac{6}{E_{\pi }^2}\right]\ ,\\
	\bar{w}_{k} & = & \frac{1}{2} k \Bigg[\frac{32 \bar{\xi} _1 \bar{W}_k'^{2}}{E_{a_0}^2 
	E_{\sigma }^4}
	+\frac{32 \bar{\xi} _1 \bar{W}_k'^{2}}{E_{a_0}^4 E_{\sigma
  	}^2}+\frac{192 \bar{\xi} _1 \bar{Y}_k^2}{E_{a_1}^6}-\frac{24 \bar{Y}_k}{E_{a_1}^4}
  	-\frac{4 
  	\left(2 \bar{\xi} _1 \bar{W}_k''+
	\bar{W}_k'\right)}{E_{\sigma }^4}-\frac{4 \left(3\bar{W}_k'+\bar{X}_k\right)}{E_{\pi }^4}
	\nonumber\\
	& & +\frac{8 \bar{\xi} _1 \bar{X}_k^2}{E_{\pi }^4 E_{\eta }^2}+\frac{8 \bar{\xi} _1 
	\bar{X}_k^2}{E_{\pi }^2 
	E_{\eta }^4}-\frac{24 \bar{Y}_k}{E_{\rho}^4}\Bigg]\ ,\\
	\bar{x}_{k} & = & \frac{1}{2} k \biggg\lbrace-\frac{32 \bar{W}_k'^{2}}{E_{a_0}^2 
	E_{\sigma }^4}-\frac{32 \bar{W}_k'^{2}}{E_{a_0}^4 E_{
	\sigma }^2}+\frac{24
  	\bar{X}_k^2}{E_{\pi }^4 E_{a_0}^2}+\frac{24 \bar{X}_k^2}{E_{\pi }^2 E_{a_0}^4}-\frac{24 
  	\bar{Y}_k}{E_{a_1}^4 \bar{\xi} _1}-
	\frac{192 \bar{Y}_k^2}{E_{a_1}^6}+\frac{4 \bar{X}_k^2}{E_{\eta
  	}^4 \left(-\bar{V}_k'+\bar{W}_k+\bar{\xi} _1 \bar{X}_k\right)}\nonumber\\
	& & -\frac{2 \bar{X}_k^2}{E_{\eta }^2 \left(\bar{V}_k'-\bar{W}_k-\bar{\xi} _1 \bar{X}_k
	\right){}^2}+\frac{4 \bar{X}_k \left(
	\bar{W}_k-\bar{V}_k'\right)}{E_{\pi }^2
  	E_{\eta }^2 \bar{\xi} _1 \left(-\bar{V}_k'+\bar{W}_k+\bar{\xi} _1 \bar{X}_k\right)}
  	+\frac{2 
  	\bar{X}_k^2}{E_{\pi }^2 
	\left(\bar{V}_k'-\bar{W}_k-\bar{\xi} _1 \bar{X}_k\right){}^2}\nonumber\\
	& & +\frac{4 \left[\bar{X}_k
  	\left(\frac{1}{\bar{\xi} _1}-\frac{\bar{X}_k}{-\bar{V}_k'+\bar{W}_k+\bar{\xi} _1 
  	\bar{X}_k}
  	\right)-3 \bar{X}_k'\right]}
	{E_{\pi }^4}+\frac{32 \left(\bar{W}_k'+\bar{\xi} _1 \bar{X}_k'+\bar{X}_k\right){}^2}
	{E_{\eta }^4
  	E_{\sigma }^2}+\frac{32 \left(\bar{W}_k'+\bar{\xi} _1 \bar{X}_k'
  	+\bar{X}_k\right){}^2}
  	{E_{\eta }^2 E_{\sigma }^4}\nonumber\\
	& & -\frac{4 \bar{X}_k}{E_{\pi }^2 E_{\eta }^2 \bar{\xi} _1}-\frac{4 \left(2
  	\bar{\xi} _1 \bar{X}_k''+\frac{\bar{X}_k}{\bar{\xi} _1}+5 \bar{X}_k'\right)}
  	{E_{\sigma }^4}
  	+\frac{24 \bar{Y}_k}
	{\bar{\xi} _1 E_{\rho }^4}\biggg\rbrace\ ,\\
	\bar{y}_{k} & = & \frac{1}{2} k \Biggg\lbrace-\frac{16 \vec{q}^{\,2} \bar{Y}_k^2}{E_{\pi 
	}^4 E_{a_1}^2 \left(\omega _n^2+\vec{q}^{\,2}
	\right)}+\frac{32 \vec{q}^{\,2} \bar{Y}_k^2}{E_{a_0}^4 E_{a_1}^2
  	\left(\omega _n^2+\vec{q}^{\,2}\right)}-\frac{16 \vec{q}^{\,2} \bar{Y}_k^2}{E_{\pi }^2 
  	E_{a_1}^4 \left(\omega _n^2
	+\vec{q}^{\,2}\right)}+\frac{32 \vec{q}^{\,2}
  	\bar{Y}_k^2}{E_{a_0}^2 E_{a_1}^4 \left(\omega _n^2+\vec{q}^{\,2}\right)}\nonumber\\
	& & +\frac{32 \vec{q}^{\,2} \left(\bar{\xi} _1 \bar{Y}_k'+\bar{Y}_k+\bar{Z}_k'\right){}^2}
	{E_{a_1}^2 E_{\sigma }^4
  	\left(\omega _n^2+\vec{q}^{\,2}\right)}+\frac{32 \vec{q}^{\,2} \left(\bar{\xi} _1 \bar{Y}
  	_k'+\bar{Y}_k+\bar{Z}_k'\right){}^2}
	{E_{a_1}^4 E_{\sigma }^2 \left(\omega
  	_n^2+\vec{q}^{\,2}\right)}-\frac{16 \vec{q}^{\,2} \bar{Y}_k^2}{E_{a_0}^4 
  	\left(\omega _n^2+
  	\vec{q}^{\,2}\right)
	\left(\bar{\xi} _1 \bar{Y}_k+\bar{Z}_k\right)}\nonumber\\
	& & +\frac{4 \bar{Y}_k \left(5 \bar{\xi}_1 \bar{Y}_k+\bar{Z}_k\right)}{E_{a_0}^4 \bar{\xi} _1 
	\left(\bar{\xi} _1 \bar{Y}_k
	+\bar{Z}_k\right)}+\frac{16 \vec{q}^{\,2} \bar{Y}_k^2}{E_{\pi }^4 E_{\rho }^2 \left(\omega
  	_n^2+\vec{q}^{\,2}\right)}+\frac{16 \vec{q}^{\,2} \bar{Y}_k^2}{E_{\pi }^2 E_{\rho }^4 
  	\left(\omega _n^2
	+\vec{q}^{\,2}\right)}\nonumber\\
	& & -\frac{16 \vec{q}^{\,2} \left[\bar{Z}_k
  	\left(\bar{\xi} _1 \bar{Y}_k'+\bar{Y}_k\right){}^2+2 \bar{Z}_k \bar{Z}_k' \left(\bar{\xi} _1 \bar{Y}
  	_k'+\bar{Y}_k\right)-\bar{\xi} _1
	\bar{Y}_k \bar{Z}_k'^{2}\right]}{E_{\sigma }^4 \bar{Z}_k 
	\left(\omega_n^2+\vec{q}^{\,2}\right) \left(\bar{\xi} _1 \bar{Y}_k+\bar{Z}_k\right)}-\frac{8 
	\bar{\xi} _1 \vec{q}^{\,2}
	\bar{Y}_k^3}{E_{\pi }^4 \bar{Z}_k \left(\omega _n^2+\vec{q}^{\,2}\right) \left(\bar{\xi} _1
  	\bar{Y}_k+\bar{Z}_k\right)}\nonumber\\
	& & -\frac{32 \vec{q}^{\,2} \bar{Z}_k'^{2}}{E_{\rho }^4 E_{\sigma }^2 \left(\omega _n^2+\vec{q}
	^{\,2}\right)}
	-\frac{32 \vec{q}^{\,2}\bar{Z}_k'^{2}}{E_{\rho }^2 E_{\sigma }^4 \left(\omega _n^2+\vec{q}^{\,
	2}\right)}
	-\frac{4 \bar{Y}_k}{E_{\eta }^4 \bar{\xi} _1}+\frac{4 \left(\frac{\bar{Y}_k}{\bar{\xi} 
	_1}+\frac{2 \bar{\xi} _1 
	\bar{Y}_k^3}{\bar{\xi}_1 \bar{Y}_k \bar{Z}_k+\bar{Z}_k^2}-3 \bar{Y}_k'\right)}{E_{\pi }^4}\nonumber\\
	& & +\frac{4 \left[-2 \bar{\xi} _1
  	\bar{Y}_k''-\frac{\bar{Y}_k}{\bar{\xi} _1}+\frac{4 \bar{Z}_k \left(\bar{\xi} _1 \bar{Y}_k'+
  	\bar{Y}_k\right){}^2
	+8 \bar{Z}_k \bar{Z}_k' \left(\bar{\xi} _1 \bar{Y}_k'+\bar{Y}_k\right)-4\bar{\xi} _1 \bar{Y}_k
  	\bar{Z}_k'^{2}}{\bar{Z}_k \left(\bar{\xi} _1 \bar{Y}_k+\bar{Z}_k\right)}-5 \bar{Y}_k'\right]}
  	{E_{\sigma }^4}
  	\Biggg\rbrace\ ,\\
	\bar{z}_{k} & = & \frac{1}{2} k \Bigg\lbrace\frac{16 \bar{\xi} _1 \vec{q}^{\,2} \bar{Y}_k^2}
	{E_{\pi }^4 E_{a_1}^2 \left(\omega _n^2
	+\vec{q}^{\,2}\right)}+\frac{16 \bar{\xi} _1 \vec{q}^{\,2} \bar{Y}_k^2}
	{E_{\pi}^2 E_{a_1}^4 \left(\omega _n^2+\vec{q}^{\,2}\right)}
	-\frac{8 \bar{Y}_k}{E_{a_0}^4}-\frac{8 \bar{\xi} _1 \vec{q}^{\,2} \bar{Y}_k^2}{E_{\pi }^4 
	\left(\omega_n^2+\vec{q}^{\,2}\right) \left(\bar{\xi} _1 \bar{Y}_k+\bar{Z}_k\right)}
	+\frac{32 \bar{\xi} _1 \vec{q}^{\,2} \bar{Z}_k'^{2}}{E_{\rho }^4 E_{\sigma }^2 
	\left(\omega_n^2+\vec{q}^{\,2}\right)}\nonumber\\
	& & +\frac{32 \bar{\xi} _1 \vec{q}^{\,2} \bar{Z}_k'^{2}}{E_{\rho }^2 E_{\sigma }^4 \left(\omega 
	_n^2+\vec{q}^{\,2}\right)}
	-\frac{16 \bar{\xi}_1 \vec{q}^{\,2} \bar{Z}_k'^{2}}{E_{\sigma }^4 \bar{Z}_k \left(\omega _n^2+\vec{q}
	^{\,2}\right)}
	+\frac{4 \left(-\frac{2 \bar{Y}_k \bar{Z}_k}{\bar{\xi} _1 \bar{Y}_k+\bar{Z}_k}-3\bar{Z}_k'\right)}{E_{\pi }
	^4}\nonumber\\
	& & -\frac{4 \left[\bar{Z}_k' \left(\bar{Z}_k-4 \bar{\xi} _1 \bar{Z}_k'\right)+2 \bar{\xi} _1 \bar{Z}_k 
	\bar{Z}_k''\right]}{E_{\sigma }^4 \bar{Z}_k}\Bigg\rbrace\ ,
\end{IEEEeqnarray}
with:
\begin{IEEEeqnarray}{rCl}
	E_{\sigma}^{2} & = & k^{2}+\omega_{n}^{2}+2\bar{V}_{k}'+4\bar{\xi}_{1}\bar{V}_{k}''\ ,\\
	E_{\pi}^{2} & = &  k^{2}+\omega_{n}^{2}+2\bar{V}_{k}'\ ,\\
	E_{a_{0}}^{2} & = & k^{2}+\omega_{n}^{2}+2\bar{W}_{k}\ ,\\
	E_{\eta}^{2} & = & k^{2}+\omega_{n}^{2}+2\bar{W}_{k}+2\bar{\xi}_{1}\bar{X}_{k}\ ,\\
	E_{\rho}^{2} & = & k^{2}+\omega_{n}^{2}+2\bar{Z}_{k}\ ,\\
	E_{a_{1}}^{2} & = & k^{2}+\omega_{n}^{2}+2\bar{Z}_{k}+2\bar{\xi}_{1}\bar{Y}_{k}\ .
\end{IEEEeqnarray}
In this case, the meson masses read:
\begin{IEEEeqnarray}{rCl}
	m_{\sigma} & = & \sqrt{2\bar{V}_{k}'+4\bar{\xi}_{1}\bar{V}_{k}''}\ ,\\
	m_{\pi} & = &  \sqrt{2\bar{V}_{k}'}\ ,\\
	m_{a_{0}} & = & \sqrt{2\bar{W}_{k}}\ ,\\
	m_{\eta} & = & \sqrt{2\bar{W}_{k}+2\bar{\xi}_{1}\bar{X}_{k}}\ ,\\
	m_{\rho} & = & \sqrt{2\bar{Z}_{k}}\ ,\\
	m_{a_{1}} & = & \sqrt{2\bar{Z}_{k}+2\bar{\xi}_{1}\bar{Y}_{k}}\ .
\end{IEEEeqnarray}
\section{Data interpolation}\label{sec:interpolation}
$n$ data points $y_{j}$ at sites $x_{j}$ are approximated by a cubic spline $f$, such that the following expression is minimized:
\begin{equation}
	p\sum_{j=1}^{n}w_{j}\left\vert y_{j}-f(x_{j})\right\vert^{2}
	+(1-p)\int \lambda(t)\left\vert D^{2}f(t)\right\vert^{2}\mathrm{d}t\ .
\end{equation}
The first term is an error measure, whereas the second a roughness measure. The default value 
for the weights $w_{j}$ as well as for the weight function $\lambda$ is one. The integration 
has to be performed over the smallest interval containing all data sites. $p$ is a smoothing parameter. This method is used via the MATLAB \texttt{csaps} function.

For both fits in Fig.~\ref{fig:full}, all weights are equal to one and $p$ is chosen to be $1\times 10^{-4}$. In Fig.~\ref{fig:anomaly}, we have $p=1\times 10^{-5}$ and $w_{1}=1\times 10^{4}$ at $x_{1}=0$.
\end{appendix}

\footnotesize{
\bibliographystyle{unsrt}}
\bibliography{mybib}
\end{document}